# ESG In Corporate Filings: An AI Perspective

Irene Aldridge[1]
Payton Martin[2]
Cornell University
This draft: November 24, 2022
First draft: June 15, 2022

## Abstract

Using Artificial Intelligence (AI) techniques, we quantitatively examine the mentions of ESG terms in the U.S. corporate filings with the SEC over the 2019-20 period. We find that in our sample of companies, the management tends to focus ESG discussions in corporate filings along three dimensions: 1) diversity, 2) hazardous materials, and 3) greenhouse gasses. However, overall, the companies and their stakeholders approach ESG from the following three "ESG pillars": 1) a combination of greenhouse gasses, data security and inclusion, 2) a tradeoff between emissions and product quality, and 3) product labeling. These dimensions can be effective ESG ratings aspects for companies.

We further measure the investor appreciation of the corporate ESG communication through a cross-sectional analysis of the ESG discussions on the contemporaneous and subsequent stock returns. Using standard linear regression methodology, we find that the markets are much more responsive to the broad spectrum of ESG pillars, as opposed to the management "party line." However, when we account for potential correlation of ESG communication dimensions using AI-based techniques, the statistical superiority of the full communication disappears and only the official corporate message carries through to the investors. We find that the strategic ESG messaging in corporate filings matters and significantly impacts corporations' forward-looking returns.

[1] iea22@cornell.edu, irene@AbleMarkets.com, irene.aldridge@gmail.com
[2] pcm65@cornell.edu
The authors are grateful to Elroy Dimson of Cambridge University Judge Business School for generous suggestions.

# 1. Introduction

The literature on ESG and U.S. corporate filings analysis is vast. In this paper, we apply a new approach to the filings' analysis: a set of Artificial Intelligence techniques to analyze the importance of ESG communication in the corporate filings. Our specific contributions are:

1. We propose a way to separate the management's strategic messaging on a given ESG topic from the overall corporate and stakeholder activity in the same area.
2. To assess the nature of the management messaging and the broader stakeholder communication on ESG in the corporate filings, we propose a methodology for identification of "ESG dimensions" which can be used as optimal data-driven ratings of companies.
3. We measure the shareholder response to the ESG discussions in corporate filings using:
   ○ Traditional OLS linear regression on the underlying terms, and
   ○ Collinearity-free analysis of the dimension-driven variables.
4. We find that ESG messaging in the corporate filings matters to investors and significantly affects corporate returns.

This paper is organized as follows: Section 2 considers the ESG debate that motivated our study. Section 3 discusses the relevant literature on the corporate filings. Section 4 tests a "base case" – a set of standard cross-sectional OLS regressions. Section 5 introduces Singular Value Decomposition: an unsupervised learning technique we will use in identification of key trends in the data. Section 6 repeats the analysis of Section 4 with the AI-augmented variables. Section 7 concludes.

# 2. ESG

## 2.1. ESG Overview

According to the Principles for Responsible Investing (PRI), a UN-supported network of investors, dedicated to promoting sustainable investment through ESG, as of 2020, over 3,000 global investors with a combined $100 trillion under management have signed a commitment to incorporate ESG in their investment approach. Fast forward to 2022, and at least 24 U.S. States, including Texas, Florida, West Virginia, North Dakota, Oklahoma, Minnesota, Idaho, South Carolina, Louisiana, Idaho, Wyoming, Arizona, Kentucky, Utah, Indiana, Missouri, Ohio and South Dakota, have either banned ESG fund investing in the State-controlled investing vehicles or are actively considering such a ban.

What contributed to such a sharp turnaround in the ESG domain? The bans on ESG are often politicized. It is a fact that most of the 24 States trying to ban ESG happen to have Republican leadership. A Bloomberg article from August 31, 2022, aptly titled "Red State Republicans' War on ESG Will Have Losses on Both Sides"[3] argued that 1) the Republicans were at war with the environment and sustainability, which may lead to the same Republicans' demise, and 2) that the Republicans "War" on ESG will hurt the investors and the environment. It has also been reported that the Republicans view ESG as a vehicle for "woke" policies.[4]

Most recently, however, anti-ESG rhetoric appears to spread in the liberal circles as well. The front-page New York Times opinion piece aptly titled "One of the Hottest Trends in the World of Investing Is a Sham"[5] argued that the contemporary ESG standards do not reflect the environmental impacts of the companies. For example, Coca–Cola and Pepsi are often highly rated on ESG, yet their products have been shown to be major contributors to diabetes of the population at large and their packaging a major cause of the world's pollution. The article went as far as to liken ESG with "regular capitalism at its slickest: ingenious marketing at the service of profits". The author further called on the Securities and Exchange Commission to step in and not only regulate the firms rated on the ESG, but examine their entire supply chain in the process of ratings as well. Oh, and in conclusion, the author called for an ESG ratings "overhaul" that "may not be as kind to corporate America."

Perhaps the biggest elephant in the room is the performance of the ESG funds. ESG funds are institutional investment vehicles that hold stocks of companies with high ESG ratings. According to Cornell (2021), average performance of ESG funds has been subpar at best. Figures 1 and 2 show Google comparisons of the S&P 500 ETF (NYSE: SPY) and one of the most popular ESG ETFs: ESGV, managed by Vanguard, at intraday and longer investment horizons. Figure 1 shows the intraday return on September 30, 2022. As the Figure 1 illustrates, Vanguard ESG ETF closely tracking, yet underperforming, SPY in 2022 YTD. Similar patterns can be seen with other ESG ETFs, for example, FlexShares STOXX US ESG Select Index Fund (BATS: ESG). The performance of the latter is a little better than that on the Vanguard ESG ETF, ESGV. However, BATS: ESG still underperforms the S&P 500 ETF, SPY.

---

---

          

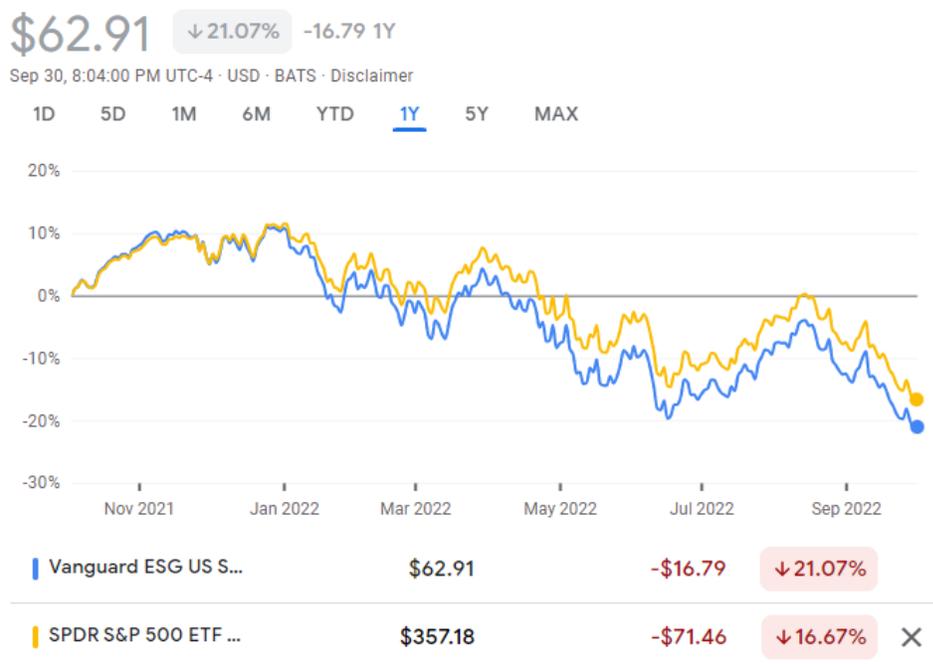

Figure 1. 2022 YTD performance of ESGV vs. SPY.

The confusion among different ESG ratings priorities potentially reflects the inherent economic difficulties in devising such metrics. As Dimson, Karakas and Li (2021) note, ESG applications are rife with structural problems, such as a free-rider problem. For example, the costs of the ESG implementation may be attributed to one group of investors, while all the investors share the benefit.

Dimson, Karakas and Li (2021) document that large investors have a strong influence on their investees' behavior with respect to environmental and social governance. Using the Collaboration Platform provided by the Principles for Responsible Investment (PRI), the researchers find higher levels of success when large investors take a hands-on approach with corporations and influence their environmental actions, rather than a passive back-seat reliance on ratings.

Recent studies on investor preferences on ESG also include Krüger, Sautner and Starks (2020). Krüger, Sautner and Starks (2020) aggregate findings from the 439 respondents to their institutional investors survey and find that the majority of the respondents to be pro-ESG. However, it is not clear if Krüger, Sautner and Starks (2020) control for the self-selection bias in such a voluntary reporting to the researchers' survey questions. Self-selection occurs when an agent (in this case, a survey respondent) is asked to voluntarily perform a certain task. The agents who are more aligned with the task will be more likely to comply as opposed to the agents who are not fit for the task. For example, when students are asked to volunteer their grades, well-performing students are more likely to report their marks than students who underperformed. Similarly, on social media, people are more likely to post when they experience events that cast them in a positive light, rather than a negative or neutral viewpoint. According to Heckman (2010), among many others, selection bias is a serious issue in statistics that arises when researchers choose what data points are collected in a survey. This can lead to an incorrect impression of how many people feel about certain issues. In the survey methodology of Krüger, Sautner and Starks (2020), it is plausible that investors who do not consider ESG to be useful did not even respond at all!

Krüger, Sautner and Starks (2020) explain that they employed professional survey designers to eliminate all bias in the survey. However, it is not clear how many prospective investors were asked to complete the survey, except that the total number was in excess of 1,018 senior investors approached by the authors through a survey service, in addition to the unspecified number of conference attendees in Paris, London and Toronto polled by Krüger, Sautner and Starks (2020). In other words, as the ratings themselves, the investor opinions about the ratings and even the studies on the latter opinions may be considered to be highly subjective.

In addition, the gathering of opinions at conferences may itself induce a bias via peer interactions. For example, such bias in a classroom setting was identified and measured by Graham (2008), who explicitly measured the effect of social interactions on children's development in various learning settings. While the conference interaction setup may be viewed as different from a school environment, both refer to the impact human communication has on the outcome behavior. In the same spirit, the conference attendees' communication could potentially induce a meaningful bias on the survey results.

This study uses advanced unbiased data science methodologies to examine investor preferences about ESG. Our contribution is along several fronts:
1) We break down the ESG umbrella into modular components and examine the relevance of each with advanced AI analysis
2) We use market data to measure investment appreciation and preference for specific ESG topics in the data science frameworks
3) Our findings are strong and parsimonious enough to be helpful to designing a standardized ESG ratings framework
4) Our findings may be useful to large and small investors alike to help guide the companies along to a truly sustainable ESG investing.

## 2.2. Conflicting Ratings

Many investors rely on ESG rating companies to provide guidance on more and less sustainable firms. The ESG ratings, however, have been in a Wild West phase that has been well documented in the academic literature. For example, Chatterjee et al. (2016) and Berg, Kölbel and Rigobon (2019) compare ESG ratings from six different rating agencies. Chatterjee et al. (2016) study KLD, Thomson-Reuters (Asset4), Calvert, Financial Times (FTSE4Good),

Dow Jones (DJSI) and Innovest. Berg, Kölbel and Rigobon (2019) examine ratings from KLD, Sustainalytics, Moody's ESG, S&P Global, Refinitiv (previously Asset4) and MSCI. Chatterjee et al. (2016) document the ratings correlation among the rating agencies to range from 31% to 53%. Berg, Kölbel and Rigobon (2019) find that the correlation between the ESG ratings assigned by the different agencies ranges from 38% to 71%. A ratings correlation of 38% translates into the following: when one agency deems a company to be ESG-compliant, another agency may assign a similar rating only 38% of the time. A lack of an organized framework is clearly demonstrated when looking at these statistics. This implies that there is no single standard for what constitutes good or bad environmental practices in this industry which can lead investors into confusion about their investment portfolios as well.

Chatterjee et al. (2016) and Delmas et al. (2013) point out that the discrepancy in ratings is at least partially driven by the diverse objectives of the rating agencies. Chatterjee et al. (2016) point out that at least some of the discrepancies among the ratings agencies are driven by the raters' domestic policy issues. According to Chatterjee et al. (2016), U.S.-based raters care more about social issues than their European counterparts. For example, "raters such as KLD give credit for products with beneficial impact on the environment, while others, like FTSE4Good, employ metrics that assess the procedures to identify and fix environmental hazards, in the spirit of the ISO 14001 management standards." (Chatterjee et al. (2016), p. 9) Kim, Wan, Wang, and Yang (2019) document that institutional shareholders are more likely to directly engage on ESG with companies located nearby, reinforcing local political influence on the ratings.

Escrig-Olmedo et al. (2019) focus on sustainability in the ESG ratings following Windolph (2011), Lozano (2015) and Ben-Eli (2018). According to Escrig-Olmedo et al. (2019), sustainability covers financial-economic, environmental and social aspects, the balance among those, the intergenerational transitions, stakeholder impacts and the overall lifecycle perspectives. Escrig-Olmedo et al. (2019) first examine how the sustainability frameworks evolved over the previous 10 years, and then consider how the ratings companies adapted the sustainability metrics in their products. Among the key findings are: 1) lack of transparency among the ESG raters on how the scores are assigned; 2) lack of standards on how a particular concept is measured; 3) questionable tradeoffs: high scores in one domain may offset very low scores in another area; 4) absence of an overall score combining performance scores along environmental, social and governance axes; 5) lack of acknowledgement of stakeholder expectations leading to lower acceptance rates. In our research, we will particularly focus on the latter point to tie the corporate actions on the environment to the investor expectations.

Billio et al. (2020) analyze ESG criteria used by different rating companies and find that the lack of common definitions leads to often conflicting ratings. In response, the calls to synchronize and regulate ESG ratings have also made their way into the academic literature. For example, Karpoff et al. (2022) insist that "The SEC should mandate disclosure of E&S-related cash flow effects, including investments that alter E&S outcomes. The SEC should require reference to ESG (or E, or S, or G) ratings to be accompanied by a description of the rating method, including factors and weights."

The precedent for governmental regulation in the ESG space has long been set in Europe. Back in 2014, the European Union (EU) created a law that requires large companies to disclose information about their social and environmental practices. The Non-Financial Reporting Directive 2014/95/EU was passed mandating large corporations across Europe make public non-financial details of what they do with respect to the ESG practices. Koundouri (2021), however, finds that such disclosure contributes little to the financial returns of the companies.

To get ahead of the regulation, at least one industry association took it up upon themselves to develop and promulgate standards for ESG. The Sustainability Accounting Standards Board (SASB), now part of the International Sustainability Standards Board (ISSB), deemed essential to break down ESG into the following standards:

1. Greenhouse Gas (GHG) Emissions
2. Air Quality
3. Energy Management
4. Water & Wastewater Management
5. Waste & Hazardous Materials Management
6. Ecological Impacts
7. Human Rights & Community Relations
8. Customer Privacy
9. Data Security
10. Access & Affordability
11. Product Quality & Safety
12. Customer Welfare
13. Selling Practices & Product Labeling
14. Labor Practices
15. Employee Health & Safety
16. Employee Engagement
17. Diversity & Inclusion
18. Business Ethics
19. Competitive Behavior
20. Management of the Legal & Regulatory Environment
21. Critical Incident Risk Management
22. Systemic Risk Management

The idea of the breakdown is to provide a granular and wholesome viewpoint of the firms' multi-faceted ESG profiles.

In this paper, we follow Schmidt's (2022) and rely on the SASB classification to streamline our analysis.

## 2.3. ESG Ratings and Investor Preferences

The studies exploring aspects of ESG ratings and their quantitative impact on investors have mostly focussed on 1) the corporate stock returns and 2) accounting variables. The results generally find that firms ranking high on ESG post higher returns (Derwal (2005), Friede et al. (2015), Ortas et al. (2015), Kumar et al. (2016), Velte (2017), Zhao et al. (2018), Brogi and Lagasio (2019), De Lucia, Pazienza and Bartlett (2020), Monasterolo and DeAngelis (2020), Koundouri et al. (2021)). Also, several studies found that firms ranking high on ESG had lower risk measured as a standard deviation of returns of beta, a regression coefficient on the market returns (Kumar et al. (2016), Giese et al. (2019), Koundouri et al. (2021)). The causality, however, in most studies is unclear: do firms with higher ESG ratings produce better returns or do firms with higher returns have extra cash to spend on ESG initiatives? This remains an open question that we are also trying to address in the current study.

Statman (2004) and Bollen (2007), on the other hand, both document that ESG firms posted negative abnormal returns for years. To attract investors, Statman (2004) and Bollen (2007) suggest that ESG companies offer investors utility from the externalities of investing by aligning with investors' beliefs. USSIF (2010) reported that institutional managers' funds choosing ESG strategies grew to $3.07 trillion from 1995 through 2010. In comparison, the institutional money under management of non-ESG strategies grew to $25.2 trillion over the same period. Renneboog, Horst, and Zhang (2008) find that investors paid a price for ESG-based strategies: the ESGs underperformed non-ESG strategies by 2% to 6% across the US, Europe and Asia-Pacific. However, Renneboog, Horst, and Zhang (2008) also found that once adjusted for risk, there was no longer significant difference between ESG and non-ESG issues. Auer and Schuhmacher (2016) find that passive investing beats ESG investing. Nofsinger and Varma (2012) find that ESG-heavy firms underperform when the markets are trending up, but outperform in bad times. Bauer, Koedijk, and Otten (2005) finds that ESG firms outperform in the U.K., but underperform in the U.S.

Yen et al. (2019) find mixed evidence of ESG performance. Landi and Sciarelli (2019) and Billio et al. (2020) find no impact of the current ESG ratings on the financial returns. Billio et al. (2020) attribute this finding to the dispersion of ratings. However, Billio et al. (2020) also find that the growth stocks are less likely to have high ESG ratings than non-growth stocks. Growth stocks tend to be short on cash, investing most of their proceeds into their growth. This furthers the hypothesis that the firms are more likely to spend on ESG when they have higher disposable money reserves, which is usually preceded by high ROE and ROA.

While the impact of the entire ESG ratings on the markets remains questionable, individual components of the ESG show promising results. Edmans (2011) finds a positive relationship between employee satisfaction and long-term stock returns. Edmans (2011) documents that American companies offering the best working conditions a risk-adjusted premium return (alpha) of 3.5% per year, 2.1% above the industry benchmark during 1984-2009. Kempf and Osthoff (2007) focus on six qualitative criteria from the KLD database: community, diversity, employee

                    

relations, environment, human rights, and product. Kempf and Osthoff (2007) use KLD data from the end of 1991 until the end of 2003, and exclude alcohol, tobacco, gambling, military, nuclear power, and firearms-related firms. The researchers find that the firms ranked high along their selected six KLD categories outperformed. Statman and Glushkov (2009) also use individual KLD categories. Over the period of 1992-2007, the authors find that the firms with high scores on community, employee relations and the environment outperformed, but so did the "exclusionary" firms such as alcohol, tobacco, gambling, military, nuclear power, and firearms-related firms. Bruder et al. (2019) finds that Governance is significantly valued by investors in Europe and North America. They also find that firms scoring high on Governance and Environment in North America and Europe tend to have lower stock volatility. They do not find any significant ESG relation in Asia-Pacific.

In our research, we focus on the U.S. firms and address the questions of returns, volatility and their causality vis-a-vis firms' ESG profiles.

# 3. Corporate Filings

## 3.1. Investor Reaction to Corporate Filings

In this study, we consider corporate filings published in the U.S. SEC EDGAR database. The filings are mandated for all public companies and follow a template, but leave significant discretion to the corporation about what data to include.

Several studies note that analysts and investors pay attention to corporate filings. Ben-Rephael, Da and Israelsen (2017), for example, find that 62% of institutional attention spikes in Bloomberg terminal usage for a specific stock coincide with Earnings Announcement dates. They find that the institutional interest is distinct from what they consider to be retail interest measured by Google searches.

Drake, Roulstone, and Thornock (2015) look at the SEC EDGAR daily activity logs and find that more hits on EDGAR on the day of an earnings announcement and the day following an earnings announcement result in faster information incorporation in prices, as evidenced by a smaller post-earnings announcement price drift. You and Zhang (2011) found that investors underreact to 10-K information.

## 3.2. Managerial Impact on Corporate Communication

It has also been shown that corporate executives like to present the corporate information in the best light possible. Donoher, Reed and Storrud-Barnes (2007), for example, document that executives with higher compensation were more likely to issue earnings restatements. Other studies linking the executive incentives and corporate disclosure include seminal Yermack (1995), Shleifer and Vishny (1997) and many others.

The alignment of incentives has prompted corporate executives to engage in a well-documented phenomenon known as "window-dressing". In window-dressing, managers may report the numbers closer to the desired numbers rather than actual numbers by utilizing a variety of accounting techniques. For more details on this topic, please see Kotomin and Winters (2006), Lakonishok, Shleifer, Thaler and Vishny (1991), McCrary (2008), Musto (1999), Ng and Wang, (2004) and Weaver, Trevino and Cochran (1999), among many others.

In ESG, the practice of window dressing has taken on a new term, "green-washing." For an extensive review of greenwashing, please see de Freitas Netto et al. (2020). In our study, we consider potential greenwashing in the context of key dimensions of managerial communication.

## 3.3. Textual Analysis of Corporate Filings

Popular metrics for textual analysis of corporate filings include sentiment, tone and readability. In this study, we use a simple keyword analysis, measuring the frequency of ESG terms defined by SASB. This framework is most similar to sentiment analysis, but is different at the same time. Studies on sentiment analysis include Antweiler and Frank (2004), Tetlock (2007), Engelberg (2008), Li (2008a and 2008b), Tetlock, Saar-Tsechansky, and Macskassy (2008) and Loughran and McDonald (2011, 2016, 2017).

## 3.4. Corporate filings and ESG

Tindall (2022) proposes that the corporate interest in the eco-friendly initiatives may be best disclosed by the corporate filings of the businesses themselves. Specifically, Tindall (2022) posits that it is shareholder activism that drives the corporate management to adopt ESG-friendly stance and actions. To determine which organizations have shareholders interested in pivoting to a more ESG-friendly corporate environment, Tindall (2022) studies corporate filings that the public U.S. corporations make with the SEC. Tindall (2022) specifically focuses on the forms documenting shareholder proposals. According to the SEC Rule 14a-8, every shareholder who owns at least $2,000 or 1% of equity can submit a proposal (500 words maximum) at least 120 days prior to the annual shareholders' meeting, in addition to 13 administrative criteria we will hereby omit for the brevity of the discussion. The proposals meeting all the acceptance criteria will be put to vote at the annual meeting and will be recorded

in the form submitted to the SEC. Tindall (2022) searches for the phrase "climate change" in forms Def14A, where the shareholder proposals are recorded.

Outside of ratings, several studies also used news and self-reported ESG disclosures as measures of corporate ESG activities. Aureli et al. (2020), for example, run an event study on the firms' publication of ESG disclosures and find that investors react significantly to the publications. Kiriu and Nozaki (2020) analyze self-reported ESG statements of corporations using a Natural Language Processing (NLP) model. Kiriu and Nozaki (2020) find that such an automated platform discerns the ESG actions better than the self-reported activities and produces more accurate ratings as a result. Borms et al. (2021) uses a corpus of Dutch news (as opposed to the self-reported ESG activities) to create ESG ratings using NLP techniques.

# 4. Analysis Methodology

## 4.1. Separating Managerial Communication from Broader Corporate Activity

The objective of our study is to determine what, if any, corporate communication in the context of ESG is favored by investors.

From a corporate communications perspective, when the management wants to promote a specific ESG message, they are likely to repeat the relevant ESG terms several times in a given filing. Thus, the relative number of mentions of a particular ESG term likely corresponds to the corporate intent to address the specific topic. The topics that are more likely to be emphasized by the management are also likely to be mentioned more extensively in the filings. To identify these terms, we use a count of different ESG terms as the proxy of managerial interest in the specific topics.

In addition to the formal corporate "party line", however, the announcements can mention other ESG topics in passing. For example, many shareholders can bring a specific ESG-related issue to a vote, and that issue will appear in the corporate filings as well. Those issues may or may not receive the many mentions in the filings, but they will appear at least once in the documents. To account for these "broader stakeholder" concerns, we use a dummy indicator variable. For each keyword $w$ and for each company $c$, the dummy variable is:

$d_{wc} = 1$, if *any* of the SEC filings by company $c$ mention keyword $w$, and

$$d_{wc} = 0 \text{ otherwise} \tag{1}$$

## 4.2. OLS Analysis

We follow Tindall (2022) in screening the corporate filings for specific keywords. Expanding on the idea, we download and programmatically examine a large number of corporate filings without discriminating based on the form type.

In the corporate filings from the SEC EDGAR database, we are looking for ESG-related keywords corresponding to the SASB categories of ESG terms: Greenhouse Gas (GHG) Emissions, Air Quality, Energy Management, Water & Wastewater Management, Waste & Hazardous Materials Management, Ecological Impacts, Human Rights & Community Relations, Customer Privacy, Data Security, Access & Affordability, Product Quality & Safety, Customer Welfare, Selling Practices & Product Labeling, Labor Practices, Employee Health & Safety, Employee Engagement, Diversity & Inclusion, Business Ethics, Competitive Behavior, Management of the Legal & Regulatory Environment, Critical Incident Risk Management and Systemic Risk Management. We programmatically search for the keywords in lower case and mark their occurrences.

After extracting the relevant count of keywords in the corporate filings, we examine whether there exist strong cross-sectional dependencies between corporate messaging on any given ESG topic and following forward-looking returns on the corporate equity. In the process, we are completely bypassing the existing ESG rating methodologies in favor of raw data coming directly from corporate communications.

## 4.3. Artificial Intelligence (AI) Analysis

Since the corporate filings are not very frequent, we take the SEC filings for the 2019-20 period to predict 2021 forward-looking cross-sectional returns for our sample of companies. The natural step at this point is to run linear regressions of the forward-looking returns on the number of mentions of different SASB keywords in the SEC filings. However, such analysis may be distorted by collinearity in data. Collinearity arises when two or more of the independent variables in a regression are correlated. Then the matrix is no longer of full rank and the results can be highly inaccurate. To solve this problem, we turn to the Singular Value Decomposition (SVD), an Artificial Intelligence technique credited to independent discoveries of Eugenio Beltrami (1835–1899) and Camille Jordan (1838–1921) (see Martin and Porter, 2012). Belsley and Klema (1974) explain how SVD solves collinearity: by eliminating the low-singular-value components from the decomposition, we are simultaneously reducing the rank of the matrix without any guesswork on our part.

While the applications of SVD to Finance are not completely novel, they are quite recent. To date, very few researchers have taken advantage of all of what SVD has to offer and applied it

 

to Financial data. Some applications include Wang (2017). For a comprehensive survey of SVD techniques and their financial applications, see Aldridge and Avellaneda (2021).

Once we divest of collinearity, we apply standard cross-sectional regressions on the data to ascertain the topics that are most (least) praised and priced in by the investors.

## 4.3.1 Why SVD?

In the OLS analyses, there is a risk that the columns are correlated. For example, we can expect that keywords "emissions" and "ghg" are correlated, simply because the terms can be considered to be close in their meaning.

When the data in the columns is indeed related, the columns are no longer linearly independent. This affects the rank of the matrix and the inverse required by the OLS equation. As a result, the OLS fails due to so-called collinearity of the columns. A traditional approach for dealing with collinearity is to remove the columns one by one and see whether the model performance improves. Such an approach, known as *dropout* in machine learning, can be useful, but also time-consuming and error prone.

Instead of the traditional dropout, we follow Belsley and Klema (1974) and construct orthogonal vectors from the original data using Singular Value Decomposition (SVD). SVD is a close cousin of the Principal Component Analysis (PCA) family of techniques that is rapidly gaining ground in Finance. A recent survey of the techniques can be found in Giglio, Kelly and Xiu (2021). Notable related techniques include IPCA by Kelly et al. (2019).

With SVD, we create orthogonal matrices $U$ and $V$ from our original matrix $X$ in the following specification:

$$X = USV^T \tag{2}$$

where the $S$ is a diagonal matrix of singular values. We focus our particular attention on the singular vector matrix, $V$.

Singular values for the dummy variables indicating whether a given company mentioned a specific term are shown in Appendix X, Figure 1. Figure 2 in the Appendix X shows singular values for the actual count of mentions table.

## 7.2.2 Optimal cut-off

The singular vectors *V* of the decomposition are orthogonal by construction. They contain the coefficients that when each is multiplied by their respective column of the original data, they produce the optimal "drivers" of the dataset (see Chamberlain and Rothschild (1983) among many others). The singular vectors corresponding to the largest singular values explain the most volatility in the dataset, and, thus, represent the most influential drivers of the data (see Bai & Ng, 2002). Including all the singular vectors leads to potential overfitting.

After the singular vectors are computed, we need to decide exactly how many of those are the most important to our analysis. Several approaches exist to determine the optimal cut-off for the number of singular vectors, including the dataset-shape based Marcenko & Pastur (1967) and explained-variance based Bai & Ng (2002).

Here, we will use the Bai & Ng (2002)-based approach to determine the optimal cut-off in the number of significant singular values. Bai & Ng (2002) propose that the optimal cut-off should be based on the trade-off between the variance value-add from each additional singular vector versus the potential overfitting that the addition of each vector entails.

The variance explained by each vector can be easily computed from the singular values, squared and normalized. As described in Aldridge and Avellaneda (2021), the singular values in *S* summarize the proportion of variability of the data explained by each corresponding singular vector:

$$\% \; variance \; explained \; = \; \left|\left| u_k s_k v_k^T \right|\right|^2 / ||X||^2 = s_k^2 / \sum_j s_j^2 \tag{3}$$

where $||.||^2$ is a Frobenius norm:

$$||A||^2 = \sum_i \sum_j \left| a_{ij} \right|^2 = Tr\left( A A^H \right) \tag{4}$$

where $Tr(.)$ is trace, the sum of the elements on the main diagonal, and $A^H$ is a conjugate transpose. For real numbers, such as the numbers in our datasets, the conjugate transpose is a simple transpose and equation (2) becomes:

$$||A||^2 = Tr\left( A A^H \right) = Tr\left( A A^T \right) \tag{5}$$

It can also be shown that trace equals the sum of the matrix eigenvalues, and that, therefore,

$$||A||^2 = Tr\left( A A^T \right) = \sum_i s_i^2 \tag{6}$$

After the optimal number of factors is selected, the optimal approximate factors are used to estimate dependencies and make predictions just like in any traditional factor model.

# 5. Data

## 5.1. SEC EDGAR Filings

We created a custom Python-based download engine to obtain text of corporate filings available on the SEC EDGAR, beginning from 2019 for each stock. Despite the automation, the process was both time-consuming and overwhelming to the Internet bandwidth. The connection was dropped by the SEC EDGAR several times over the download period, most likely due to the considerable bandwidth usage of our application. After a certain time, we were no longer able to connect to the SEC EDGAR using our custom app, apparently after being temporarily or permanently blacklisted by the SEC EDGAR servers.

From November 15, 2021, through December 3, 2021 (the last day our app connected to the SEC servers), we were able to download full historical filings for 109 stock symbols,[6] resulting in the download of 17,727 corporate filings. The filings were loaded alphabetically by the stock symbol (ticker) from 'AA'. We were unable to connect to the SEC EDGAR database after the download of the filings for 'BRSP' was completed. Despite the connection issue, we strongly believe that our sample is 1) randomly drawn from the stock population without any bias toward a specific industry or any other dimension, and 2) is sufficiently large to analyze in our experiment.

In the filings data, we observe several interesting trends that are worth mentioning. We segregate the number of mentions across each of the SASB category by:
1) Filing year (2019, 2020 or 2021)
2) SEC EDGAR Form Type
3) Company industry
4) State of incorporation

Selected highlights of the ESG mentions in the filings are shown in the sections 6.1.1-6.1.4. In our analysis, we use the ESG mentions in two ways: 1) as a "dummy variable", a binary indicator of whether or not the term was mentioned in the company's filings, and 2) the actual count of the mentions in the given firm's filings.

It is natural for some of the ESG mentions variables to be correlated. For example, companies discussing greenhouse gasses may also talk about emissions. Appendixes 1 and 2 show

---

[6] The tickers included in the analysis were: AZPN, AZZ, BAC, BA, BANC, BAND, BANF, BANR, BATRA, BATRK, BAX, BBBY, BBIO, BBSI, BBW, BBWI, BBY, BCAB, BCBP, BCC, BC, BCEL, BCML, BCO, BCOR, BCOV, BCPC, BCRX, BDC, BDN, BDSI, BDTX, BDX, BEAM, BECN, BE, BELFB, BEN, BEPC, BERY, BFAM, BFC, BFIN, BFLY, BFS, BFST, BGCP, BG, BGFV, BGS, BGSF, BHB, BH, BHE, BHF, BHLB, BHVN, BIGC, BIG, BIIB, BILL, BIO, BIPC, BJ, BJRI, BK, BKD, BKE, BKH, BKI, BKNG, BKR, BKU, BLBD, BL, BLD, BLDR, BLFS, BLI, BLKB, BLK, BLMN, BLNK, BLUE, BLX, BMI, BMRC, BMRN, BMTX, BMY, BNFT, BNGO, BOH, BOKF, BOOM, BOOT, BOX, BPMC, BPOP, BPTH, BRBR, BRBS, BRC, BRG, BRKL, BRKR, BRMK, BRO, BRP, BRSP

---

correlations of occurrences of the mentions in both dummy variable form and the actual count form. As the Tables show, companies likely to mention "wastewater" have a 46% and 47% chance of also mentioning "emissions" and "product quality", respectively. As expected, companies discussing "emissions" are also likely to talk about "greenhouse" and "air quality". Overall, however, the ESG mentions correlations are reasonably low.

### 5.1.1. SEC EDGAR Filings by Filing Year

Greenhouse gas emissions (keyword 'ghg') were mentioned more often in 2019 than in 2021. However, the use of keyword "greenhouse" rose from 2019 to 2021 and the use of keyword "emissions" stayed relatively flat, as Figure 2 shows. The keywords "greenhouse", "ghg" and "emissions" may potentially create collinear structure in the data, which we dealt with using SVD analysis as described in the Methodology and Results sections. "Air quality" became a concern in 2020, likely coinciding with the COVID-induced lockdowns, as shown in Figure 3.

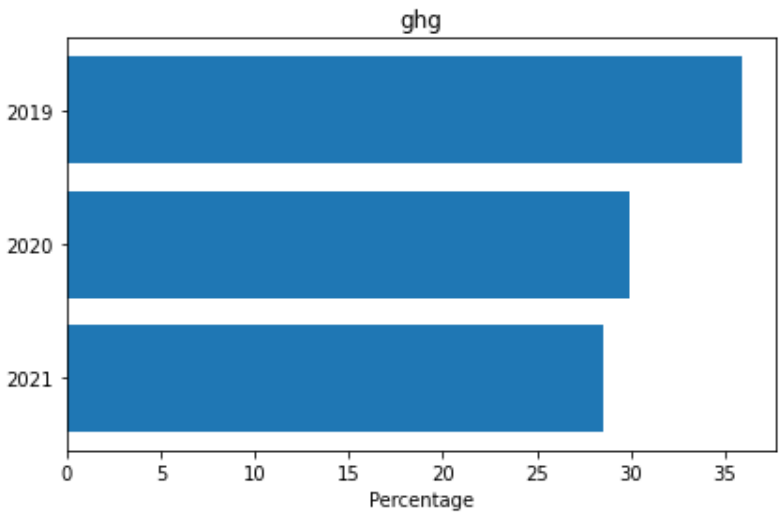

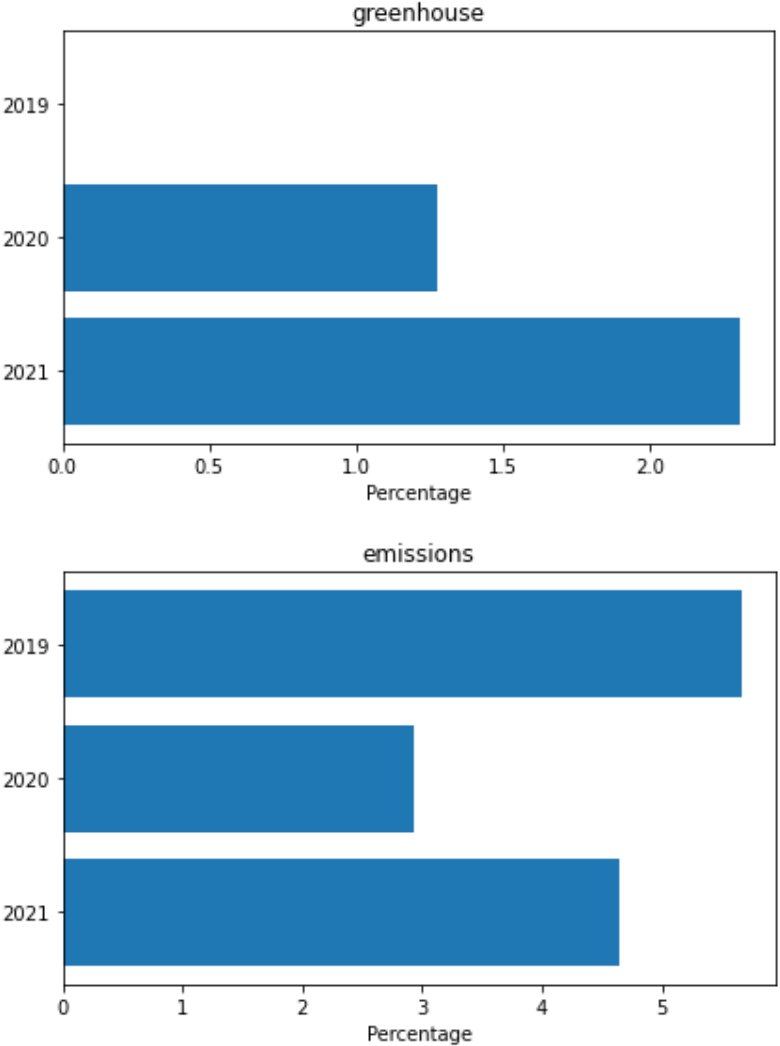

Figure 2. Mentions of the greenhouse gas emissions in our sample of the SEC EDGAR corporate filings.

     

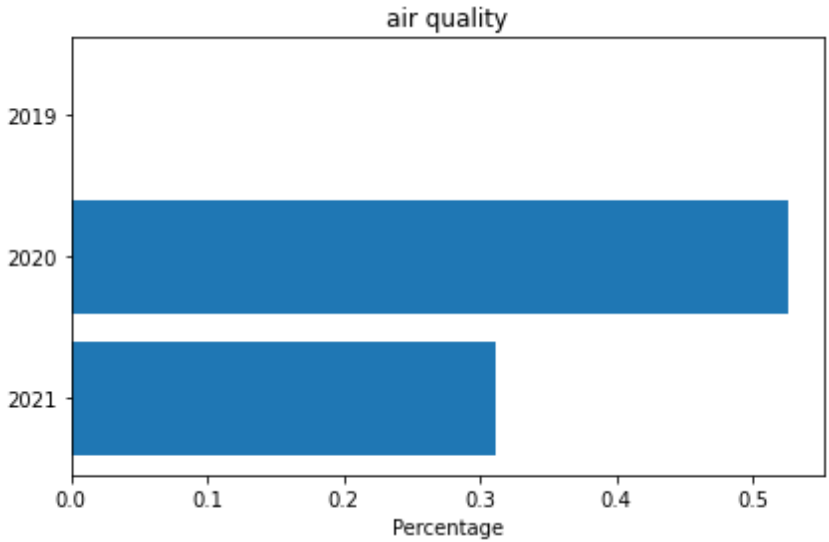

Figure 3. The prevalence of the "Air quality" keyword as a percentage of the filings in our sample that mention the keyword.

### 5.1.2. SEC EDGAR Filings by Form Type

Some keywords appear in more forms than others. For example, 'ghg' (for 'greenhouse gas') appears on almost all the form types, as shown in Figure 4. On the other hand, "product safety" appears in a handful of forms: Def14A and DefA14A, a summary of shareholder proposals and additional proxy materials, as well as 424B5, a pre-IPO supplementary prospectus, 10-Q, a quarterly report mandated by the SEC, and 8-K, a form notifying shareholders of an unscheduled material event (Figure 5).

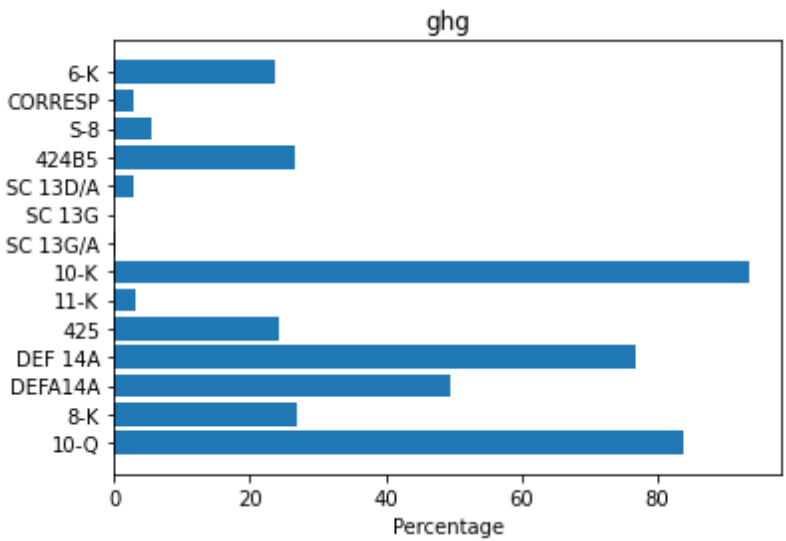

Figure 4. 'Ghg', an acronym for "greenhouse gas", distribution on various SEC forms.

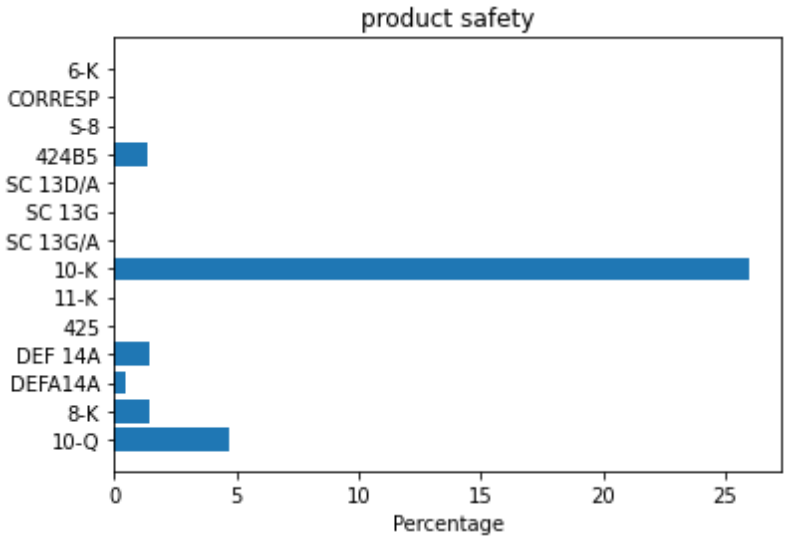

Figure 5. "Product safety" distribution on various SEC forms.

### 5.1.3. SEC EDGAR Filings by Company Industry Sector

We also broke down our sample of corporate filings by the industry sector. Using the standard industry sector classification, 'emissions' occurred at least once in reports across most industries. Surgical and Medical Instruments and Apparatus mentioned 'emissions' most often, followed by Retail-Eating places, Laboratory Analytical Instruments and Biological Products, as shown in Figure 6.

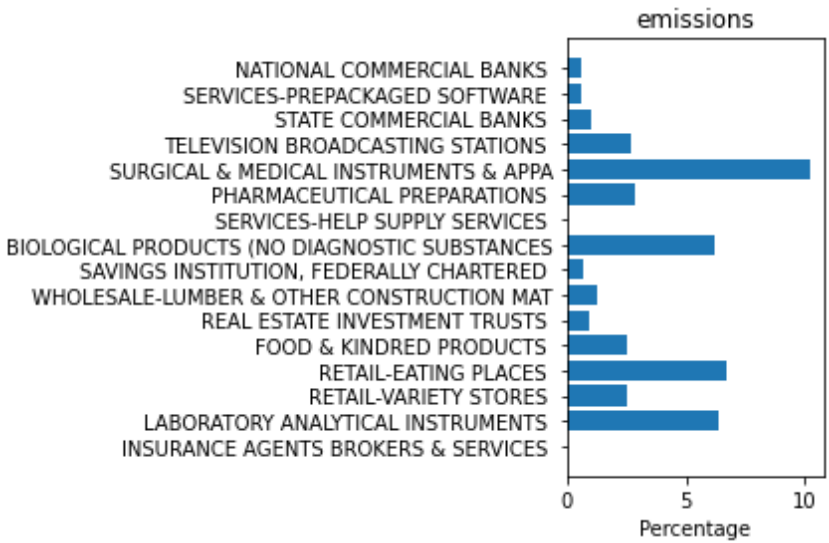

Figure 6. Mentions of "emissions" by industry.

"Air quality" was a concern to anyone working away from home or dealing with tenants during COVID: State Commercial Banks, Surgical and Medical Equipment, Real Estate Investment Trusts, Food & Kindred products, Retail Eating Places, Real Variety Stores and so on, as shown in Figure 7. Curiously, "product quality" was only of interest to Insurance Agents, Brokers and Services, at least in our sample (Figure 8).

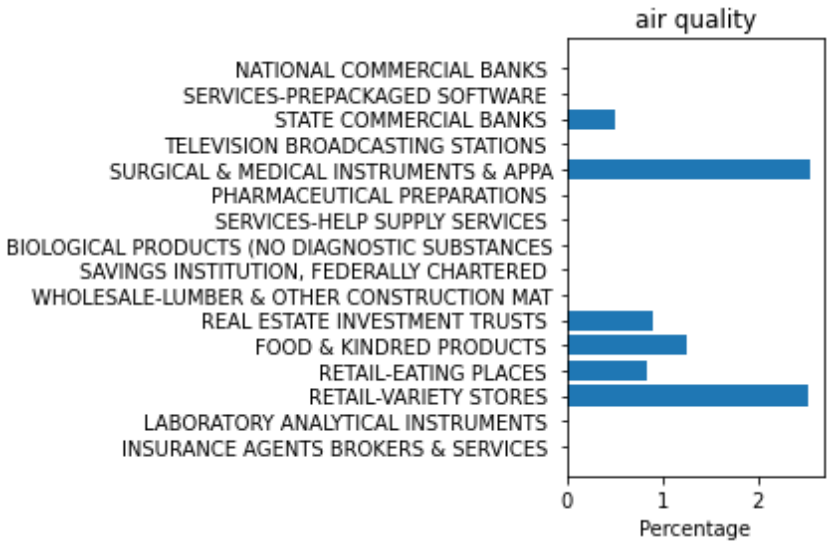

Figure 7. Mentions of "air quality" by industry.

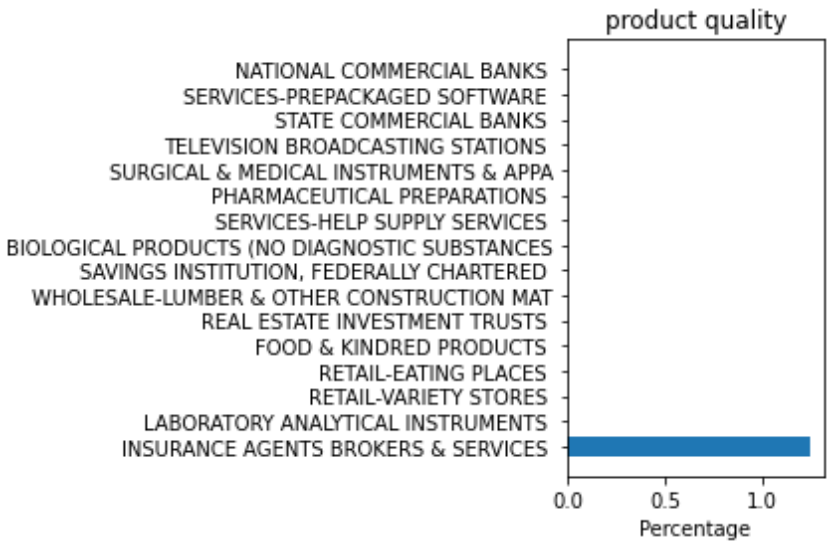

Figure 8. Mentions of "product quality" by industry.

                20

### 5.1.4. SEC EDGAR Filings by State of Incorporation

Our sample of firms was incorporated in only a handful of states: FL, IN, WI, VA, NJ, NY, OK, CA, MD, TX and DE. Companies domiciled in NJ, CA, MD and DE were concerned about 'air quality.' (Figure 9). Inclusion, however, was top-of-the-mind in all the states in the sample, as Figure 10 shows.

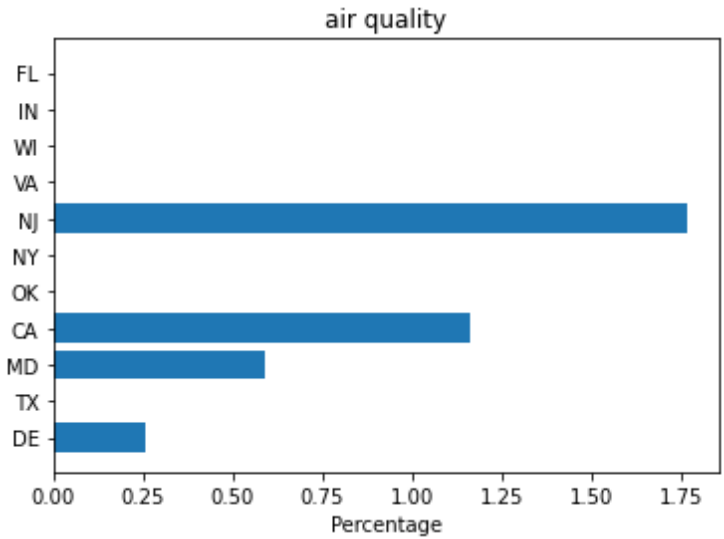

Figure 9. Mentions of "air quality" by state.

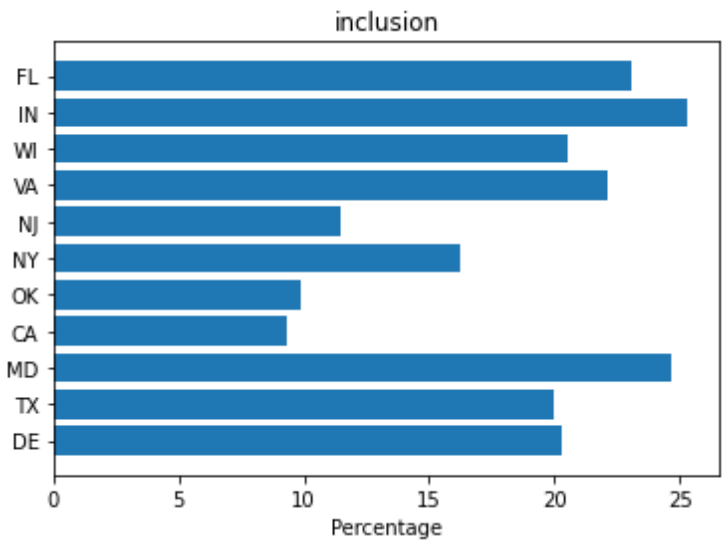

Figure 10. Mentions of "inclusion" by state.

"Employee safety" mattered most in Texas, Indiana and Wisconsin (Figure 11), while "employee engagement" was highlighted in Wisconsin and California (Figure 12).

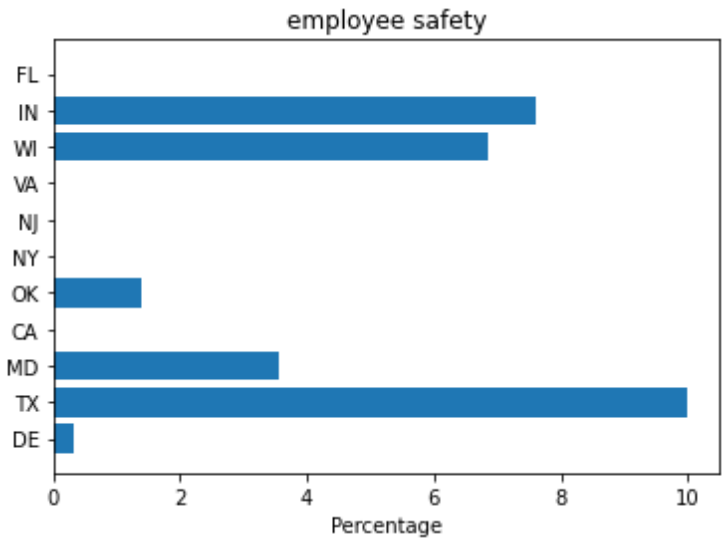

Figure 11. Mentions of "employee safety" by state.

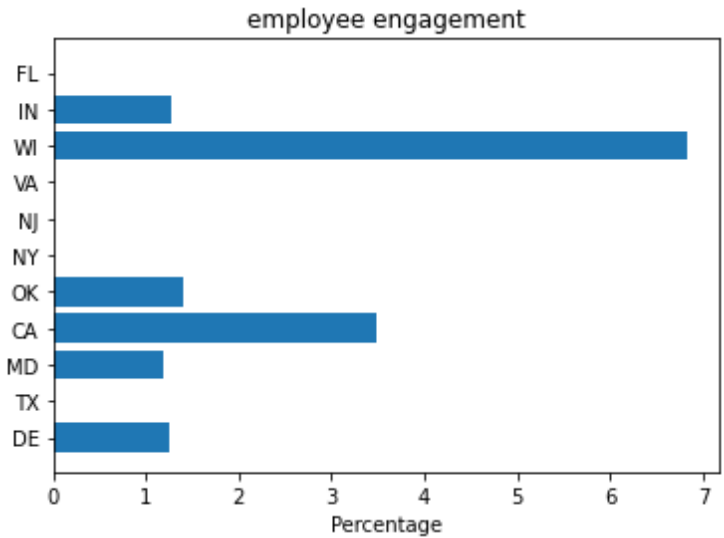

Figure 12. Mentions of "employee engagement" by state.

We can expect the mentions data to be correlated. For example, when a company management talks about "greenhouse", the company is also likely to talk about "emissions." Actual correlations for the data used in the analysis are presented in Appendices. As shown in the tables, firms talking about "greenhouse" are indeed 49% likely to talk about "emissions" as well, on average. This is just one example of a pair of highly correlated data series in the data.

## 5.2. Stock Price Information

All the stock price information was downloaded from Yahoo! Finance. We used Adjusted Close prices for each stock to measure the in-sample 2020-21 return and the 2021 stock return as the forward-looking dependent variable in our analysis.

All returns were computed as simple daily returns: $r_{i,t} = P_{i,t}/P_{i,t-1} - 1$, where $P_{i,t}$ is the Adj Close price for stock $i$ on day $t$ as reported by Yahoo! Finance. Next, we computed simple averages, standard deviations, skewness and kurtosis for all stocks to use in the cross-sectional analysis. The summary statistics on those for 2019-20 and 2021 periods are reported in Tables 1 and 2, respectively.

The Tables show interesting dynamics among the stock characteristics. For example, as expected, the distribution of the skewness of returns was much more positive in the 2019-20 period than it was in 2021. Similarly, the distribution of kurtosis was much more positive in the 2019-20 than in 2021. However, even in 2021 the average levels of kurtosis were above 8 (down from 15 in 2019-20). Such high levels indicate extremely "fat tails" in the distribution of returns and present extreme risks, or "black swan" events (Taleb (2010)), not captured by the standard deviation. We consider all of these variables in the cross-sectional analysis.

TABLE 1. Summary Cross-Sectional Return Statistics for the Stocks Used in the Analysis for the 2019-20 period.

| Metric | E[R] | Std[R] | Skewness[R] | Kurtosis[R] |
|--------|------|--------|-------------|-------------|
| Count | 109 | 109 | 109 | 109 |
| Mean | 0.172% | 3.933% | 0.513 | 15.190 |
| Std | 0.221% | 2.412% | 2.110 | 34.889 |
| Min | -0.138% | 1.356% | -3.901 | -0.040 |
| 1% | -0.089% | 1.744% | -1.333 | 0.676 |
| 3% | -0.025% | 1.858% | -1.053 | 1.513 |
| 5% | 0.000% | 1.976% | -0.942 | 1.739 |
| 10% | 0.017% | 2.380% | -0.675 | 3.626 |
| 25% | 0.053% | 2.808% | -0.295 | 6.714 |
| 50% | 0.097% | 3.375% | 0.183 | 9.068 |
| 75% | 0.201% | 4.252% | 0.600 | 12.677 |
| 90% | 0.416% | 5.477% | 1.696 | 19.905 |
| 95% | 0.683% | 6.988% | 2.340 | 33.488 |
| 97% | 0.749% | 7.760% | 3.515 | 55.609 |
| 99% | 0.916% | 16.019% | 9.269 | 110.193 |
| max | 1.310% | 20.402% | 17.088 | 346.791 |

TABLE 2. Summary Cross-Sectional Return Statistics for the Stocks Used in the Analysis for the 2019-20 period.

| Metric | E[R] | Std[R] | Skewness[R] | Kurtosis[R] |
|---|---|---|---|---|
| Count | 109 | 109 | 109 | 109 |
| Mean | 0.083% | 2.764% | 0.469 | 8.486 |
| Std | 0.186% | 1.474% | 1.924 | 22.996 |
| Min | -0.612% | 1.157% | -7.288 | -0.327 |
| 1% | -0.504% | 1.165% | -3.176 | 0.009 |
| 3% | -0.302% | 1.232% | -1.359 | 0.236 |
| 5% | -0.280% | 1.374% | -0.915 | 0.315 |
| 10% | -0.108% | 1.546% | -0.403 | 0.468 |
| 25% | 0.026% | 1.803% | -0.123 | 1.134 |
| 50% | 0.099% | 2.242% | 0.225 | 2.143 |
| 75% | 0.166% | 3.272% | 0.659 | 5.865 |
| 90% | 0.248% | 4.973% | 1.513 | 15.264 |
| 95% | 0.344% | 5.713% | 2.176 | 23.325 |
| 97% | 0.425% | 6.155% | 2.537 | 75.090 |
| 99% | 0.483% | 7.439% | 10.319 | 139.953 |
| max | 0.780% | 9.017% | 10.898 | 150.632 |

# 6. Results

## 6.1. OLS Regressions

As a first step in the analysis, we run cross-sectional multivariate OLS regressions of both contemporaneous 2019-2020 returns and forward-looking 2021 returns per company on the mentions of various ESG topics in their respective 2019-2020 filings. We are seeking to ascertain whether some, and if so, which of the topics matter more to investors and are, therefore, reflected in the concurrent and subsequent corporate returns.

We conduct a total of sixteen regressions, eight each with 1) a regression on the dummy variable indicating whether a given keyword was mentioned at all in a particular corporation's filings in 2019-2020, and 2) the actual number of mentions of the given keyword in each corporation's 2019-2020 filings.

For each keyword $w$ and for each company $c$, the dummy variable is:
$d_{wc} = 1$ , if *any* of the SEC filings by company $c$ mention keyword $w$, and
$d_{wc} = 0$ otherwise                                                                      (1)

The dependent variables ($y$) change:
1) from 2019-2020 concurrent period to forward-looking 2021 period to examine potential causality, and
2) from the cross-sectional average daily returns E[R] to the standard deviation of the returns Std[R] to Skewness and Kurtosis.

### 6.1.1. Regressions of the "Dummy" indicator of ESG term mentions

The detailed results of the dummy variable regressions of the existence of specific ESG mentions in 2019-20 results are presented in Appendix III. Here, Table 3 summarizes the key findings that are statistically-significant at the 90%+ confidence level.

TABLE 3. Summary of OLS regression results of various return characteristics on the binary indicator of mentions of ESG terms ("dummy variable") by corporations in the SEC filings over 2019-20. The relevant ESG keywords were proposed by SASB. The regressions are conducted contemporaneously on the 2019-20 data. T-statistics are shown in parentheses. Coefficients statistically-significant at 90% level are shown in bold; *** indicates 99% statistical significance.

|  | E[R] 2019/20 | Std[R] 2019/20 | Skew[R] 2019/20 | Kurt[R] 2019/20 |
|---|---|---|---|---|
| greenhouse | -0.0015 (-1.613) | -0.0132 (-1.332) | -8.4362 (-0.639) | -0.7531 (-0.823) |
| emissions | **0.0029 (4.045)*** | **0.0180 (2.313)** | 6.4279 (0.622) | 0.8584 (1.197) |
| ghg | 0.0003 (0.402) | 0.0128 (1.561) | **19.7731 (1.822)** | 0.7043 (0.936) |
| air quality | **0.0034 (2.839)** | **0.0280 (2.151)** | 8.2862 (0.480) | 1.4799 (1.237) |
| wastewater | **-0.0027 (-2.396)** | -0.0190 (-1.520) | -0.5713 (-0.034) | -0.2422 (-0.211) |
| hazardous materials | -0.0006 (-1.171) | 0.0040 (0.703) | 2.5952 (0.343) | 0.2160 (0.412) |
| human rights | 0.0002 (0.299) | -0.0093 (-1.169) | **-17.5332 (-1.669)** | -0.9869 (-1.354) |
| data security | **0.0011 (2.002)** | 0.0022 (0.367) | 4.3123 (0.536) | 0.5856 (1.049) |
| access and affordability | 0.0010 (0.499) | -0.0016 (-0.073) | -8.9266 (-0.299) | -0.6934 (-0.335) |
| product quality | 0.0009 (1.422) | 0.0018 (0.268) | -13.5189 (-1.543) | -0.3252 (-0.535) |
| product safety | -0.0009 (-1.052) | -0.0075 (-0.801) | **-23.7969 (-1.916)** | -1.1816 (-1.372) |
| selling practices | -0.0019 (-0.697) | -0.0242 (-0.801) | -56.8796 (-1.420) | 0.2135 (0.077) |
| product labeling | **0.0028 (2.599)** | **0.0315 (2.684)** | **71.9263 (4.615)*** ** | **2.0816 (1.926)** |
| labor practices | **0.0020 (1.802)** | **0.0481 (4.034)*** ** | **61.4690 (3.885)*** ** | **3.4257 (3.122)*** ** |
| employee health | -0.0003 (-0.347) | -0.0077 (-0.819) | -14.5514 (-1.661) | -0.7459 (-0.858) |
| employee safety | -0.0010 (-1.144) | 0.0016 (0.177) | **25.7111 (2.115)** | 1.1079 (1.314) |
| diversity | -0.0008 (-1.645) | **-0.0104 (-1.853)** | -11.8431 (0.117) | -0.4520 (-0.871) |
| inclusion | -0.0002 (-0.323) | -0.0062 (-0.799) | -12.6297 (-1.230) | -0.6822 (-0.958) |
| employee engagement | **-0.0014 (-1.749)** | -0.0111 (-1.265) | -10.1094 (-0.864) | **-1.4592 (-1.799)** |
| business ethics | -0.0002 (-0.186) | -0.0068 (-0.685) | -0.3622 (-0.028) | -0.9649 (-1.061) |
| competitive behavior | -0.0017 (-0.761) | -0.0002 (-0.008) | 18.8161 (0.567) | 1.2722 (0.533) |
| **Intercept** | **0.0015 (3.855)*** ** | **0.0342 (8.254)*** ** | **11.6408 (2.118)*** ** | 0.4533 (1.189) |
| **Model Adj. R2** | 21.8% | 23.6% | 13.4% | 34.2% |

Table 3 links the existence of corporate mentions of specific ESG factors and average returns for the same time period and shows several interesting trends. First of all, the Adjusted $R^2$ of the model is 21.8%, a very large number for a financial model explaining returns, even in a contemporaneous setting. From this $R^2$ value alone we are tempted to conclude that ESG indeed matters to investors.

Several ESG coefficients are highly statistically significant at 90%+ confidence level, supporting our findings that at least some ESG criteria are important to investors. Specifically, average returns for 2019-20 were higher for companies that mentioned "emissions", "air quality", "data security", "product labeling" and "labor practices." At the same time, the average daily 2019-20 returns for companies mentioning "wastewater" and "employee engagement" were lower than those for companies that kept silent on the topics, as summarized in Table 3.

We also note that in the results of Table 3 the intercept is positive and statistically significant. Since the markets overall went up considerably during 2019-20, we expected the intercept to be positive. The fact that the intercept is statistically significant at 99.99% statistical confidence indicates that, naturally, other factors outside of the cross-sectional stock returns can help explain by their ESG attitudes as contained in the SEC filings. Such factors may be contemporaneous market returns and the like.

As Table 3 shows, concurrent volatility was lower for companies mentioning diversity, but higher for firms talking about emissions, air quality, product labeling and labor practices. We also consider the skewness of the returns distribution. Many companies had return time series disrupted during the peak of COVID, and the resulting average returns metric may not adequately capture the quality of returns. Skewness, on the other hand, measures the proportion of the returns that were either positive or negative. Thus, a positive relationship of term's mentions on the skewness may show a more nuanced impact of ESG.

In contemporaneous 2019-20 data, companies that mentioned "product labeling" and "labor practices" had a higher number of positive daily returns over the same period. At the same time, companies talking about "employee engagement" had a higher share of negative returns in 2019-20. Once again, causality in this case is difficult to establish from the current analysis alone: it is possible that the companies with low employee engagement suffered the consequences as a result of such.

To fine-tune our analysis of risk in the cross sectional regressions on the existence of mentions further, we next regress firms' kurtosis on the dummy variables. Kurtosis of returns represents the "fattiness" of tails of the returns distribution, which translates into the risk of extreme changes. The contemporaneous and forward-looking results are shown in Table 3.

In the contemporaneous analysis of kurtosis on the existence of mentions, the Adjusted R2 is 34.2%, the highest yet of our regressions. Mentions of "product labeling" and "labor practices" were associated with the highest kurtosis stocks at the 99.999% statistical confidence level. It is possible that these companies suffered the most during COVID. Other mentions with positive contemporaneous relationships with kurtosis were "employee safety" and "ghg". "Human rights", on the other hand, was associated with a lower contemporaneous kurtosis. These results are summarized in Table 3.

                    

Moving to the forward-looking return predictability, we next examine Table 4. The Table summarizes regression results of average daily returns computed over the 2019-20 time period on the same dummy variables. The Adjusted $R^2$ is now lower at 6.2%, which is expected as we are dealing with forward-looking predictability. Examining the individual coefficients, we find that the coefficients corresponding to the following keywords are 90%+ statistically significant to the following year's stock returns and deliver a positive impact on the next year's average returns: "labor practices" and "employee health". This may be a special situation related to COVID emergencies, as reflected in the shareholders' opinions. In addition, "data security" and "product labeling" keywords had 90%+ statistically-significant, but negative relationships with later forward-looking returns. Positive and negative ESG factors are summarized in Table 4.

Table 4 also shows forward-looking dependencies of the existence of mentions on the 2021 volatility. The results were quite consistent: companies that mentioned "emissions", "air quality" and "labor practices" in 2019-20 had higher volatility in 2021 than companies that did not. At the same time, companies mentioning "diversity" in 2019-20 once again showed lower volatility in 2021. It is possible that the causality is reversed and that lower volatility companies are more likely to talk about diversity, while high-volatility companies may be more concerned about emissions, air quality and labor practices. On skewness, companies with 2019-20 mentions of "emissions", "ghg" and "wastewater" in 2019-20 had more positive return days in 2021.

Table 4 also shows the only statistically-significant ESG term that was mentioned in 2019-20 and was linked with 2021 kurtosis: "emissions". Firms mentioning emissions in 2019-20 had higher kurtosis in 2021.

TABLE 4. Summary of OLS regression results of various return characteristics on the binary indicator of mentions of ESG terms ("dummy variable") by corporations in the SEC filings over 2019-20. The relevant ESG keywords were proposed by SASB. The regressions link the 2019-20 ESG mentions with the forward-looking returns computed over 2021. T-statistics are shown in parentheses. Coefficients statistically-significant at 90% level are shown in bold; *** indicates 99% statistical significance.

| | E[R] 2021 | Std[R] 2021 | Skew[R] 2021 | Kurt[R] 2021 |
|---|---|---|---|---|
| greenhouse | -0.0001 (-0.123) | -0.0065 (-0.956) | -14.2310 (-1.304) | -0.8834 (-0.999) |
| emissions | 0.0009 (1.333) | **0.0118 (2.224)** | **14.6993 (1.720)** | **1.6755 (2.419)** |
| ghg | 0.0008 (1.088) | 0.0032 (0.568) | -4.0117 (-0.447) | **1.4972 (2.059)** |
| air quality | -0.0009 (-0.835) | **0.0160 (1.815)** | -6.7775 (-0.475) | -0.8681 (-0.751) |
| wastewater | 0.0012 (1.164) | -0.0140 (-1.649) | 10.2761 (0.749) | **2.0081 (1.807)** |
| hazardous materials | -0.0003 (-0.545) | 0.0036 (0.938) | 2.8656 (0.458) | 0.5070 (1.001) |
| human rights | 0.0001 (0.154) | -0.0041 (-0.772) | -5.5528 (-0.639) | -0.7851 (-1.116) |
| data security | **-0.0012 (-2.333)** | 0.0025 (0.611) | -4.8228 (-0.725) | -0.7328 (-1.359) |
| access and affordability | 0.0007 (0.343) | 0.0004 (0.024) | -4.6777 (-0.190) | -0.1655 (-0.083) |
| product quality | -0.0008 (-1.353) | 0.0058 (1.291) | 11.1556 (1.540) | -0.7537 (-1.284) |
| product safety | 0.0002 (0.262) | 0.0044 (0.688) | 2.4159 (0.235) | -0.2772 (-0.333) |
| selling practices | 0.0024 (0.934) | -0.0023 (-0.111) | -15.2597 (-0.461) | -1.4148 (-0.527) |
| product labeling | **-0.0026 (-2.605)** | 0.0043 (0.542) | -6.5047 (-0.505) | -1.4834 (-1.421) |
| labor practices | **0.0028 (2.799)** | **0.0158 (1.953)** | 8.0018 (0.612) | 1.5666 (1.478) |
| employee health | **0.0015 (1.901)** | -0.0042 (-0.651) | -8.9236 (-0.861) | -0.5886 (-0.701) |
| employee safety | -0.0007 (-0.879) | -0.0040 (-0.635) | -9.8327 (-0.978) | -0.9413 (-1.156) |
| diversity | 0.0005 (1.045) | **-0.0094 (-2.463)** | -5.6333 (-0.911) | 0.0681 (0.136) |
| inclusion | -0.0009 (-1.329) | -0.0009 (-0.177) | 10.8516 (1.279) | -0.7672 (-1.116) |
| employee engagement | -0.0001 (-0.142) | -0.0073 (-1.213) | -11.5893 (-1.199) | -1.0074 (-1.286) |
| business ethics | 0.0000  (0.082) | -0.0039 (-0.584) | -3.6417 (-0.336) | 0.2832 (0.322) |
| competitive behavior | -0.0006 (-0.292) | -0.0019 (-0.109) | 7.8492 (0.286) | 1.5999 (0.719) |
| **Intercept** | **0.0009 (2.664)** | 0.0264 (9.384)*** | 6.5626 (1.444) | 0.0451 (0.122) |

 

| Model Adj. R2 | 6.2% | 6.3% | -3.5% | 2.8% |
|---|---|---|---|---|

## 6.1.2. OLS regressions of the actual count of mentions of ESG terms

Separately from the binary indicators of whether companies did or did not mention specific ESG terms, we examine the actual count of the mentions to ascertain whether talking a lot about a particular ESG issue has an impact on the firm's contemporaneous and future returns. The details of the cross-sectional linear regressions of the 2019-20 mentions and 1) 2019-20 returns and 2) 2021 returns are presented in Tables 5 and 6, respectively.

TABLE 5. Summary of OLS regression results of various return characteristics on the number of actual mentions of ESG terms by corporations in the SEC filings over 2019-20. The relevant ESG keywords were proposed by SASB. The regressions are conducted contemporaneously on the 2019-20 data. T-statistics are shown in parentheses. Coefficients statistically-significant at 90% level are shown in bold; *** indicates 99% statistical significance.

| | E[R] 2019/20 | Std[R] 2019/20 | Skew[R] 2019/20 | Kurt[R] 2019/20 |
|---|---|---|---|---|
| greenhouse | -0.0002 (-0.975) | -0.0005 (-0.185) | -1.9021 (-0.467) | -0.0430 (-0.177) |
| emissions | 0.0001 (1.051) | 0.0007 (0.476) | 1.1566 (0.536) | 0.0134 (0.104) |
| ghg | -0.0000 (-0.452) | **-0.0001 (-1.668)** | -0.1604 (-1.545) | **-0.0106 (-1.701)** |
| air quality | 0.0001 (0.183) | 0.0034 (0.528) | -3.9841 (-0.412) | 0.1414 (0.245) |
| wastewater | 0.0000 (0.263) | 0.0010 (0.931) | 2.5200 (1.526) | 0.1496 (1.517) |
| hazardous materials | 0.0000 (1.400) | 0.0003 (0.866) | -0.0153 (-0.031) | 0.0060 (0.204) |
| human rights | -0.0003 (-0.534) | **-0.0104 (-2.011)** | **-16.5717 (-2.148)** | **-1.0009 (-2.172)** |
| data security | 0.0000 (1.289) | 0.0008 (1.351) | 0.0074 (0.008) | 0.0569 (1.064) |
| access and affordability | -0.0006 (-0.462) | -0.0101 (-0.765) | -3.5203 (-0.178) | -0.7039 (-0.597) |
| product quality | -0.0000 (-0.340) | -0.0015 (-0.873) | -4.0162 (-1.569) | -0.1614 (-1.055) |
| product safety | -0.0000 (-0.028) | 0.0000 (0.001) | -0.2603 (-0.099) | -0.0591 (-0.375) |
| selling practices | 0.0000 (0.065) | -0.0034 (-0.223) | 3.7432 (0.164) | 1.0986 (0.808) |
| product labeling | -0.0000 (-0.380) | 0.0009 (0.479) | 2.4243 (0.905) | 0.0714 (0.446) |
| labor practices | 0.0008 (1.224) | **0.0220 (3.106)***** | **30.7770 (2.922)** | **1.7092 (2.717)** |
| employee health | -0.0000 (-0.007) | -0.0046 (-0.805) | -4.9180 (-0.581) | -0.2568 (-0.508) |
| employee safety | -0.0000 (-0.200) | 0.0027 (0.672) | **11.5151 (1.914)** | 0.5386 (1.498) |
| diversity | -0.0000 (-1.004) | **-0.0001 (-1.794)** | -0.0495 (-0.474) | -0.0024 (-0.380) |
| inclusion | 0.0000 (1.376) | 0.0000 (0.186) | -0.2726 (-0.561) | -0.0120 (-0.413) |
| employee engagement | -0.0001 (-0.904) | -0.0006 (-0.500) | 0.5391 (0.281) | -0.0987 (-0.861) |
| business ethics | -0.0000 (-0.142) | -0.0014 (-0.501) | 2.0190 (0.477) | -0.3121 (-1.235) |
| competitive behavior | -0.0003 (-0.330) | 0.0031 (0.309) | 9.5859 (0.637) | 0.5982 (0.666) |
| **Intercept** | **0.0014 (4.388)*** | **0.0404 (11.477)*** | **20.7945 (3.973)*** | **0.8715 (2.788)** |
| **Model Adj. R2** | -2.8% | 1.6% | -6.4% | -3.9% |

As Table 5 shows, there were no statistically-significant contemporaneous dependencies between the mention counts and the average returns. At the same time, as shown in Table 6, the mention counts for several ESG terms had 90%+ statistical confidence with expected returns in 2021. The term with a significant positive impact on average daily 2021 returns was "employee health": the more times the companies mentioned that in their regulatory filings during 2019-2020 were more likely to see more positive average returns in 2021. At the same time, "data security" and "product labeling" showed a significantly negative relationship with 2021 average returns, as summarized in Table 5. These results are consistent with the results of the dummy regressions shown in Table 3.

Next, we regress the mention counts on the standard deviations of returns. The full contemporaneous results are shown in Table 5, while Table 6 displays regressions of the look-ahead standard deviation data. Like with the dummy variable analysis, more frequent talk about "labor practices" resulted in higher contemporaneous and future volatility of returns. At the same time, more concern about "diversity" was linked with lower contemporaneous and future volatility. Mare mentions of "ghg" and "human rights" were linked with higher contemporaneous, but not future, volatility.

Next, we once again examined cross-sectional contemporaneous and forward-looking skewness of returns as a function of actual mentions of individual ESG terms by public corporations in the U.S. SEC filings over the 2019-20 period. Table 5 shows the statistically-significant results from the contemporaneous analysis. As Table 5 shows, "labor practices" repeated in the filings corresponded to higher contemporaneous skewness, while high mentions of "ghg" and "human rights" were associated with lower contemporaneous skewness, all at the 90%+ statistical confidence levels. The forward-looking 2021 skewness of

 

returns showed no statistically-significant dependencies on the 2019-2020 mention counts of any ESG terms, as documented in Table 6.

Kurtosis measures the risk of extreme returns. Tables 5 and 6 summarize the statistically-significant details of kurtosis regressions for the contemporaneous and look-ahead variables, respectively. As with Dummy regressions, "employee safety" and "labor practices" had a positive relationship with kurtosis and "human rights" had a negative link with a given firm's stock returns kurtosis in the contemporaneous data. In the forward-looking analysis, higher count of "inclusion" mentions in 2019-20 resulted in higher stock kurtosis in 2021. A quick Google search on the differences between diversity and inclusion came up with 'Diversity is the "what"; inclusion is the "how."' from Rita Mitjans, ADP's chief diversity and social responsibility officer.[7] In data terms, "diversity" is the actual headcount separated by diversity variables, such as gender or race, whereas "inclusion" is a measure of cultural efforts to integrate diversity into the corporate culture.

TABLE 6. Summary of OLS regression results of various return characteristics on the number of actual mentions of ESG terms by corporations in the SEC filings over 2019-20. The relevant ESG keywords were proposed by SASB. The regressions link the 2019-20 ESG mentions with the forward-looking returns computed over 2021. T-statistics are shown in parentheses. Coefficients statistically-significant at 90% level are shown in bold; *** indicates 99% statistical significance.

| | E[R] 2021 | Std[R] 2021 | Skew[R] 2021 | Kurt[R] 2021 |
|---|---|---|---|---|
| greenhouse | -0.0001 (-0.529) | -0.0008 (-0.474) | -1.2227 (-0.437) | -0.2476 (-1.056) |
| emissions | 0.0000 (0.515) | 0.0006 (0.654) | 0.4772 (0.322) | 0.1529 (1.232) |
| ghg | 0.0000 (1.325) | **-0.0000 (-1.880)** | -0.0263 (-0.369) | 0.0005 (0.092) |
| air quality | -0.0003 (-0.592) | 0.0017 (0.421) | -1.2836 (-0.193) | -0.3295 (-0.593) |
| wastewater | 0.0000 (0.734) | 0.0002 (0.306) | 0.4375 (0.386) | 0.1030 (1.085) |
| hazardous materials | -0.0000 (-0.192) | 0.0002 (1.090) | -0.2126 (-0.628) | -0.0026 (-0.093) |
| human rights | -0.0003 (-0.673) | -0.0043 (-1.334) | -2.6442 (-0.499) | -0.4821 (-1.086) |
| data security | **-0.0000 (-1.699)** | 0.0004 (1.171) | -0.3355 (-0.545) | -0.0088 (-0.170) |
| access and affordability | 0.0005 (0.522) | -0.0055 (-0.668) | -2.7797 (-0.205) | -0.0493 (-0.043) |
| product quality | -0.0002 (-1.232) | 0.0004 (0.371) | 0.5938 (0.338) | -0.2144 (-1.456) |

---

[7]https://www.adp.com/spark/articles/2019/03/diversity-and-inclusion-whats-the-difference-and-how-can-we-ensure-both.aspx#:~:text=Mitjans%3A%20Diversity%20is%20the%20%22what,that%20enables%20diversity%20to%20thrive.

| | | | | |
|---|---|---|---|---|
| product safety | 0.0000 (0.608) | 0.0006 (0.575) | -0.0222 (-0.012) | -0.0043 (-0.028) |
| selling practices | 0.0012 (0.995) | -0.0076 (-0.813) | -6.5576 (-0.420) | -0.0295 (-0.023) |
| product labeling | **-0.0003 (-2.346)** | -0.0001 (-0.104) | 0.1920 (0.104) | -0.1158 (-0.751) |
| labor practices | 0.0005 (0.858) | **0.0078 (1.800)** | -0.3922 (-0.054) | 0.2267 (0.374) |
| employee health | **0.0009 (1.989)** | -0.0015 (-0.427) | -2.6681 (-0.459) | 0.0298 (0.061) |
| employee safety | 0.0002 (0.639) | -0.0019 (-0.755) | -3.4867 (-0.844) | 0.2401 (0.694) |
| diversity | -0.0000 (-0.781) | **-0.0000 (-2.145)** | -0.0362 (-0.504) | -0.0017 (-0.290) |
| inclusion | -0.0000 (-0.268) | 0.0001 (0.567) | **0.5871 (1.757)** | 0.0130 (0.463) |
| employee engagement | -0.0000 (-0.557) | -0.0006 (-0.711) | -1.3307 (-1.009) | -0.1128 (-1.022) |
| business ethics | 0.0000 (0.038) | -0.0011 (-0.632) | -0.4857 (-0.167) | 0.0448 (0.184) |
| competitive behavior | -0.0008 (-1.062) | 0.0003 (0.052) | -0.8138 (-0.079) | -0.0261 (-0.030) |
| **Intercept** | **0.0009 (3.343)** | **0.0279 (12.905)*** ** | **7.9526 (2.212)** | **0.5566 (1.849)** |
| **Model Adj. R2** | 1.0% | 1.2% | -15.6% | -15.9% |

### 6.1.3. OLS Results Discussion

Overall, several ESG terms emerged as perennial or at least frequent winners in the OLS regressions. The existence of mentions (i.e., dummy variables) has produced a lot more statistically-significant values than did the total number of mentions. As shown in Table 1, companies talking about emissions had higher contemporaneous returns and skewness, but also higher next-year volatility and kurtosis of returns. Mentions of labor practices corresponded to the increased returns and their volatility. Mentions of diversity appeared to help reduce volatility both contemporaneously and a year later. On the other hand, companies talking most about ghg, diversity and human rights experienced lower returns and also lower volatility, as Table 2 shows. Focussing a lot on labor practices, however, delivered higher returns, but also higher volatility.

The Adjusted $R^2$ in the regressions is notable in its own right. The regressions on the dummy variables had outstanding Adjusted $R^2$, both with contemporaneous and forward-looking returns. The Adjusted $R^2$ of the regressions on the actual mention counts had very weak Adjusted $R^2$ results. Such discrepancy suggests that repeated mentions of the same topic do not impress the investors.

## 6.2. SVD Results

The singular values for the decompositions of both the indicators and the counts of ESG mentions are shown in Figures 13 and 14.

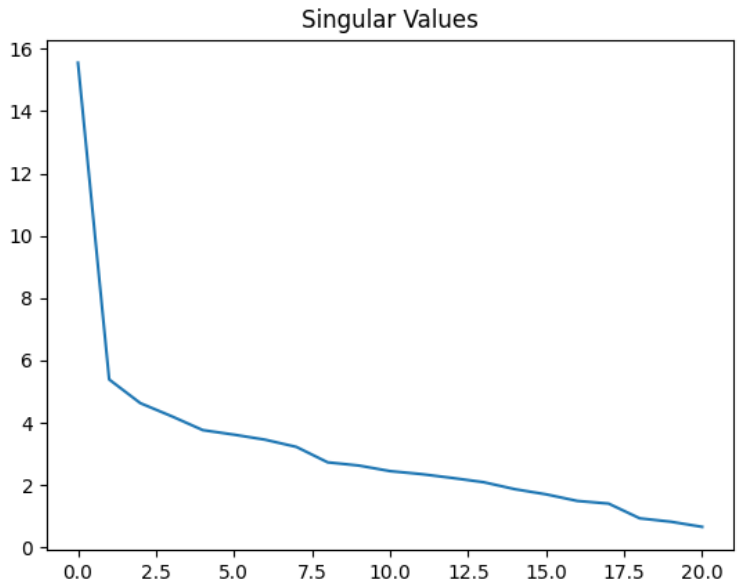

Figure 13. Scree Plot of Singular Values for the Indicators of Mentions of ESG Terms (dummy variables)

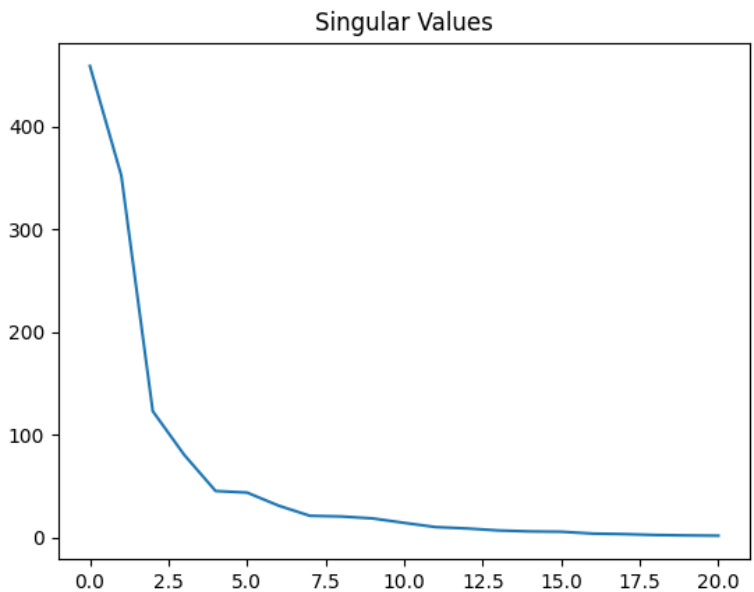

Figure 14. Scree Plot of Singular Values for the Actual Count of Mentions of ESG Terms..

For our datasets, the explained variance of the singular vectors for the dummy indicators of existence of mentions was:

var_explained = [0.592 0.071 0.053 0.044 0.035 0.032 0.029 0.026 0.018 0.017 0.015 0.014 0.012 0.011 0.009 0.007 0.006 0.005 0.002 0.002 0.001]

For the actual number of mentions of each ESG term, the explained variance was:

var_explained = [0.581 0.342 0.042 0.018 0.006 0.005 0.003 0.001 0.001 0.001 0.001 0. 0. 0. 0. 0. 0. 0. 0. 0. ]

In both datasets, the first singular vector explains nearly 60% of the entire dataset variation. In the case of the dummy variables, the explanatory power drops dramatically right after the first vector with the second singular vector explaining just 7% of the variation. For the actual count of mentions, however, the second singular vector explained over 34% of the dataset variation. The explanatory power of third and all the following singular vectors for both datasets was 5% and below. With this in mind, we choose to retain and analyze just the first three singular vectors in each dataset, setting $k^* = 3$.

### 6.2.3. The composition of the singular vectors

Next, we consider the composition of each singular vector with the objective to establish the most significant constituents corresponding to the features in the original dataset $X$, following Aldridge and Guan (2021): The feature coefficients comprising the first three singular vectors are shown in Table 7 for the dummy dataset and Table 8 for the actual count of mentions. As Table 7 shows, the largest constituents of the first and most significant singular vector V[0] are 'inclusion', 'data security', and 'selling practices". The second singular vector V[1] is dominated by "emissions" and "product quality", and the third vector V[2] is all about 'greenhouse.'

For the actual counts, the main drivers of the top 3 singular vectors are more clear-cut. The first singular vector, V[0], comprises -85% of "selling practices" and is bound to be dominated by that variable. The second singular vector, V[1], is -80% of "hazardous materials" and the third vector, V[2], is -96% "greenhouse". The detailed results are displayed in Table 8.

TABLE 7. Importance of individual features in the dummies dataset.

| Feature | V[0] | V[1] | V[2] |
|---------|------|------|------|
| **greenhouse** | -8.94252009e-02 | -0.15659949 | **-5.61108555e-01** |
| **emissions** | -2.17373238e-01 | **-0.42921843** | 3.35729540e-01 |
| **ghg** | **2.63516103e-01** | 0.26159934 | 1.92127278e-01 |
| air quality | -1.61996415e-01 | -0.35622719 | 8.67368120e-02 |
| wastewater | 2.28617201e-01 | 0.07339038 | 1.12228177e-01 |
| **hazardous materials** | **-2.85184903e-01** | 0.16292204 | 9.25400589e-02 |
| human rights | 1.35733657e-01 | -0.02983079 | 2.03404379e-02 |
| **data security** | **4.07517714e-01** | 0.28627701 | 1.25268719e-04 |
| access and affordability | -1.23936954e-02 | -0.0050507 | -1.66462461e-02 |
| **product quality** | 1.92604310e-01 | **-0.55112325** | -9.49609649e-02 |
| product safety | -1.08602672e-01 | 0.01626755 | 2.27629770e-01 |
| **selling practices** | **3.77233794e-01** | -0.10473422 | **3.71282065e-01** |
| product labeling | -2.12111301e-01 | 0.10968108 | 2.62557214e-01 |
| **labor practices** | 3.75208356e-02 | -0.1632348 | **3.75697149e-01** |
| **employee health** | -4.76215684e-05 | **0.24290418** | -1.95656295e-02 |
| employee safety | 1.36672713e-01 | -0.06402103 | -3.05426602e-01 |
| **diversity** | **-2.54584951e-01** | 0.13088774 | 4.03950515e-02 |
| **inclusion** | **4.53684422e-01** | -0.20836277 | -4.29866449e-02 |
| employee engagement | 3.88310569e-02 | -0.04592309 | -8.88355942e-03 |
| business ethics | -4.89649367e-02 | 0.03481965 | -7.54611123e-03 |
| competitive behavior | 1.02471687e-03 | -0.02238506 | -1.02676229e-02 |

TABLE 8. Importance of individual features in the actual mentions counts data.

| Feature | V[0] | V[1] | V[2] |
|---|---|---|---|
| **greenhouse** | -0.00371723 | -0.01404679 | **-9.55304407e-01** |
| emissions | 0.00096478 | 0.0038085 | 2.65236116e-01 |
| ghg | -0.02851086 | -0.08891122 | 1.27163209e-01 |
| air quality | 0.04247134 | 0.18143769 | 2.58418014e-02 |
| wastewater | -0.16157496 | -0.42099855 | -2.36898838e-03 |
| **hazardous materials** | -0.28629099 | **-0.79644712** | 1.09264234e-02 |
| human rights | 0.09807468 | 0.10562503 | 9.62973093e-04 |
| data security | 0.13822581 | 0.02638994 | 5.68667906e-03 |
| access and affordability | -0.16657476 | -0.01976084 | 3.18624653e-03 |
| product quality | -0.06389057 | 0.00981075 | -6.23528036e-04 |
| product safety | 0.16395867 | -0.00894301 | 2.62267588e-03 |
| **selling practices** | **-0.85325075** | 0.32875505 | 1.26589968e-05 |
| product labeling | 0.01596003 | 0.00125799 | 8.08769259e-04 |
| labor practices | 0.10843421 | -0.05342201 | 4.79237184e-03 |
| employee health | 0.07764077 | -0.01009384 | 1.19713027e-03 |
| employee safety | -0.08479241 | 0.01080112 | 8.37615808e-05 |
| diversity | 0.19371484 | -0.14465648 | -6.24492171e-04 |
| inclusion | 0.07436485 | -0.04493181 | 1.95079220e-03 |
| employee engagement | 0.07441314 | -0.03152359 | 1.99269375e-04 |
| business ethics | -0.00092516 | 0.00102123 | -2.66267359e-05 |
| competitive behavior | 0.00134713 | -0.00099854 | 1.76835915e-04 |

## 6.2.4. Approximate Factor Computation

To verify that the examination of the singular vectors does not result in spurious explanation, we compute the optimal factors $F_k$ following the POET methodology developed by Fan, Liao and Mincheva (2011):

$$F_k = XV[k]/\sqrt{s_k} \tag{7}$$

The approximate factors, also known as eigenfactors, comprise the optimal factors for a given data set (for proof, see Aldridge and Avellaneda (2021), among others).

Following the estimation of the optimal number of factors in the preceding section, we compute the factors $F_k$ for $k$ = [0, 1, 2], and then:

1) Verify that the $F_k$ are indeed highly dependent on the variables highlighted in Tables 7 and 8.
2) Verify that the factors $F_k$ indeed have a high explanatory power in their respective datasets.

## 6.2.5. Verifying Approximate Factor Meaning

To verify the composition of factors shown in Tables 7 and 8, we conduct simple correlation analysis of the optimal approximate factors on the underlying data columns. The correlation analysis in explaining the optimal factor structure has been used by Muravyev, Vasquez and Wong (2018) among others.

Tables 9 and 10 summarize the correlations of the first three optimal approximate factors with various columns of the dummy variables (Table 9) and the actual mention counts (Table 10).

Comparing Tables 7 and 9, we note that the results are consistent. For example, the most significant components of V[0] used to construct F0 identified in Table 7 were:
- Ghg (26.3%)
- Hazardous materials (-28.6%)
- Data security (-40.7%)
- Selling practices (-37.7%)
- Diversity (-25.5%)
- Inclusion (45.4%)

 

In Table 9, the variables most highly correlated with F0 are:
- Ghg (59.5%)
- Data security (60.1%)
- Inclusion (66.7%)

The correlation of F0 with "diversity" dummy is 19.9% and with "hazardous materials" dummy is only 0.6%. The "hazardous materials" and "diversity" likely dropped off our implied dependencies in the first optimal factor because of the variables' collinearity with other variables in the dataset. Overall, however, "ghg", "data security" and "inclusion" showed strong correlations, although "data security" dummy is now positively correlated with F0, while it was a negative component of V[0] as shown in Table 7. The correlation coefficients do not exactly match the coefficients in V[0] because of the correlation in the underlying data. While V[0] are constructed perfectly orthogonal, the resulting factors are a product of V[0] with the correlated underlying data and will exhibit slightly different dependencies.

Following the methodology of Muravyev, Vasquez and Wong (2018), we can also attempt to intuitively interpret this set of factors that comprise the most significant driver of the mentions dummies dataset. The most important composite factor comprises ghg, data security and inclusion as the three key pillars of the ESG representation of the SASB factors, as reflected in the corporate filings.

Similarly, for factor F1, the most significant components of V[1] underlying F1 identified in Table 7 were:
- Emissions (-42.9%)
- Product quality (-55.1%)
- Data security (28.6%)
- Employee health (24.3%)

In Table 9, the variables most highly correlated with F1 are:
- Emissions (-59.7%)
- Product quality (-61.0%)
- Data security (26.3%)

The correlation of employee health with F1 came out to 16.6%, somewhat lower than we expected. Here, the initial composition of F1 reflects very closely F1 correlation with the underlying data. An intuitive interpretation of this second pillar of ESG would be that it comprises "data security", "emissions" and "product quality".

Finally, for factor F2, the most significant components of V[2] underlying F2 in Table 7 were:
- Greenhouse (-56.1%)
- Emissions (33.6%)
- Selling practices (37.1%)
- Labor practices (37.6%)
- Employee safety (-30.5%)

                    

In Table 9, the variables most highly correlated with F2 are:
- Ghg (44.0%)
- Data security (44.1%)
- Product safety (40.8%)
- Product labeling (58.4%)

The correlation coefficients of F2 with the underlying data do not match the V[2] components, but do reflect the coefficients of F0 and F1.

TABLE 9. Correlations of the optimal factors with the individual underlying data time series (features) in the dummies dataset.

| Feature | F[0] | F[1] | F[2] |
|---|---|---|---|
| greenhouse | -12.8% | -44.3% | -22.9% |
| **emissions** | -7.4% | **-59.7%** | 32.3% |
| **ghg** | **59.5%** | 15.1% | **44.0%** |
| air quality | -9.9% | -41.1% | -11.2% |
| **wastewater** | 27.9% | -31.9% | 25.8% |
| hazardous materials | 0.6% | 2.35% | 41.2% |
| human rights | 34.5% | 3.41% | 4.4% |
| **data security** | **60.1%** | 26.3% | **44.1%** |
| access and affordability | 3.4% | 16.3% | 3.2% |
| **product quality** | 30.5% | **-61.0%** | 21.3% |
| **product safety** | 5.2% | 10.2% | **40.8%** |
| selling practices | 20.1% | -13.9% | 39.7% |
| **product labeling** | -0.6% | 3.2% | **58.4%** |
| labor practices | 19.2% | 7.4% | 28.7% |
| employee health | 17.2% | 16.2% | -10.0% |
| employee safety | 18.5% | -8.6% | -6.9% |
| diversity | 19.9% | 5.8% | 39.0% |
| inclusion | **66.7%** | 1.0% | 36.2% |

| employee engagement | 26.9% | -14.2% | 21.3% |
| business ethics | 14.7% | -13.5% | 26.4% |
| competitive behavior | 21.1% | 6.66% | -1.9% |

TABLE 10. Correlation of the optimal factors with the individual underlying data time series (features)  in the actual mentions counts data.

| Feature | F0 | F1 | F2 |
| --- | --- | --- | --- |
| greenhouse | -7.8% | -8.4% | -18.3% |
| emissions | -15.1% | -18.6% | -2.2% |
| **ghg** | -11.4% | -34.8% | **97.4%** |
| air quality | -14.2% | -20.7% | 8.4% |
| wastewater | -7.9% | -10.7% | -1.6% |
| **hazardous materials** | -17.2% | **-80.6%** | 6.4% |
| human rights | -3.4% | 2.7% | 8.2% |
| data security | 7.8% | -7.0% | 7.8% |
| access and affordability | 0.8% | -1.7% | 2.8% |
| product quality | -7.2% | -11.7% | 9.2% |
| product safety | 6.3% | 0.1% | 6.2% |
| selling practices | -1.3% | 1.4% | 4.1% |
| product labeling | -10.5% | -30.6% | 4.2% |
| labor practices | -0.7% | -7.3% | 30.3% |
| employee health | -4.3% | -1.3% | -5.9% |
| employee safety | -4.6% | -0.7% | 13.7% |
| **diversity** | **91.4%** | **-62.6%** | 2.8% |
| inclusion | 7.8% | -32.8% | 9.9% |
| employee engagement | 12.7% | -2.9% | 0.9% |

| business ethics | 3.7% | -1.7% | 10.9% |
| competitive behavior | 1.1% | 3.1% | 0.9% |

Table 10 displays correlation results of the optimal approximate factors based on the actual count of mentions in the corporate filings. The factors were constructed using the components of the singular vectors shown in Table 8. As Table 10 shows, factor F0 is almost perfectly correlated with mentions of "diversity" (correlation of 91.4%), factor F1 is negatively correlated with "hazardous materials" (-80.6%) and "diversity" mentions (-62.6%) and factor F2 is nearly perfectly positively correlated with "ghg" (97.4%).

The factor composition delivers a unique insight into the preferences of 1) management (Table 10) and 2) shareholders and other stakeholders (Table 9). The number of times an ESG term is mentioned in the corporate filings is at a full management discretion and is aligned with the management's strategies. As discussed in the introduction, the shareholders' opinions can also be registered in the corporate filings if they submit proposals for a vote at the annual meeting. While their message may be drowned in the larger corporate communication machine, their voice will still be heard in the dummy variable data measuring whether a particular topic was mentioned at all, as opposed to the number of times the topic appeared in the corporate filings. Thus, while the factors in Table 10 represent the management's perspective, the factors in Table 9 show key concerns of the rank-and-file shareholders.

The dichotomy between the two is interesting. For one, the management's voice is very focussed across all the companies. The first and dominant factor is singularly focussed on diversity. The second factor is mostly about hazardous materials with a continuous dash of diversity. The third management factor is strictly about greenhouse gasses.

The shareholders' key trends, however, are more diverse. The first factor representing the dominant concerts of shareholders comprises "greenhouse gasses", "data security" and "inclusion", a practical integration of diversity into the corporate culture. The second factor can be interpreted as a tradeoff between "greenhouse emissions" and "product quality". The last factor in our analysis is dominated by "product labeling".

## 6.2.5. Stock Returns vis-a-vis Optimal Factors

To empirically consider the optimal factor influence on the returns, we run cross-sectional OLS Linear Regressions of market returns characteristics on the approximate factors. In essence, we repeat our cross-sectional regression analysis from section 6.1 using approximate factors as our independent variable instead of the raw dummies and mention count datasets.

It may indeed seem spurious that the gist of the explanatory variables dataset produces the same most significant values as the linear regression with a dependent variable, cross-sectional stock returns. However, this example illustrates the true power of AI: the key drivers dominating a given dataset will drive variability in other dependent variables as well. Tables 11 - 14 below summarize results of OLS Linear Regressions of various return characteristics on the three optimal approximate factors.

As Tables 11-14 show, using the optimal approximate factors leads to a significant increase in the explanatory power of the actual mention count models, but not in the dummy variable models. For example, the Adjusted $R^2$ of the regression of contemporaneous average returns on the actual mention count in the filings went up from -2.8% in Table 2 to 2.2% in Table 14. While the dummy regression results in Tables 11 and 12 are muted compared with Table 3 and 4, the dummy regression outcomes in Tables 1 and 2 may include considerable collinearity distorting the output, while the results in Table 11 and 12 are constructed to be nearly collinearity-free.

For the dummy variables, an increase in the second optimal factor F1 is shown to dampen contemporaneous returns, as shown in Table 11. An increase in the third factor F2 raises the contemporaneous returns instead, as well as resulting in increases in the contemporaneous standard deviation, skewness and kurtosis. As shown in Table 9, the second dummy factor F1 is negatively correlated with "emissions" (correlation of -59.7%) and "product quality" (correlation of -61.0%). We can surmise that discussions of emissions and product quality are actually linked to increases in contemporaneous returns. The third optimal dummy factor F2 is most correlated with "product labeling" (correlation of +58.4%), so the shareholder interest in correctly-labeled products helps boost contemporaneous returns, along with their volatility, skewness and kurtosis.

Table 12 shows look-ahead results for the dummy optimal factors. As the coefficients show, no dummy factors are potent in explaining the average 2021 returns. However, the first optimal dummy factor F0 is shown to reduce forward-looking volatility. As documented in Table 9, the first optimal factor includes mostly "greenhouse gasses", "data security" and "inclusion", so the increased shareholder interest in these reduces future volatility.

The "management ESG factors" hold up well in both contemporaneous and look-ahead regressions, as shown in Tables 13 and 14. The first factor F0 is linked with lower contemporaneous returns. F0 also appears to lower contemporaneous and forward-looking volatility. As shown in Table 10, F0 is dominated by mentions of "diversity" (correlation of +91.4%). Thus, higher management attention to diversity correlates with lower contemporaneous returns, but also lower contemporaneous and future return volatility.

The second optimal management ESG factor F1 is also shown to reduce contemporaneous returns. Per Table 10, F1 is 1) negatively correlated with the number of mentions of "hazardous materials" (correlation of -80.6%), and 2) less importantly, negatively correlated with "diversity"

 

(correlation of -62.6%). Thus, F1 offsets the negative effect of diversity on the contemporaneous returns, but amplifies the negative influence on the forward-looking returns. Likewise, the mentions of "hazardous materials" are positively related to the contemporaneous returns, but negatively to the forward-looking returns.

The third optimal management factor F2 is also positively significant for the forward-looking returns, yet negatively impacts the forward-looking volatility, as shown in Table 14. As documented in Table 10, the optimal management factor F2 is +97.4% correlated with the number of mentions for "greenhouse gasses". Thus, extensive mentions of "greenhouse gasses" in 2019-20 corporate returns appear to have helped companies raise 2021 returns and lower 2021 volatility.

TABLE 11. Summary of regression results of various return characteristics on optimal approximate factors of the binary indicator of mentions of ESG terms ("dummy variable") by corporations in the SEC filings over 2019-20. The relevant ESG keywords were proposed by SASB. The regressions are conducted contemporaneously on the 2019-20 data. T-statistics are shown in parentheses. Coefficients statistically-significant at 90% level are shown in bold; *** indicates 99% statistical significance.

| | E[R] 2019/20 | Std[R] 2019/20 | Skew[R] 2019/20 | Kurt[R] 2019/20 |
|---|---|---|---|---|
| F0 | -0.0001 (-0.810) | -0.0022 (-1.371) | -0.1187 (-0.844) | -2.5581 (-1.122) |
| F1 | **-0.0007 (-2.777)** | -0.0027 (-0.948) | 0.0076 (0.029) | 4.5183 (1.080) |
| F2 | **0.0012 (2.955)** | **0.0143 (3.118)*** | **0.7682 (1.875)** | **16.5442 (2.490)** |
| **Intercept** | **0.0015 (4.599)*** | **0.0373 (10.196)*** | 0.4030 (1.237) | **12.2982 (2.329)** |
| **Model Adj. R2** | 9.9% | 6.6% | 0.7% | 4.6% |

TABLE 12. Summary of regression results of various return characteristics on optimal approximate factors of the binary indicator of mentions of ESG terms ("dummy variable") by corporations in the SEC filings over 2019-20. The relevant ESG keywords were proposed by SASB. The regressions are conducted on forward-looking returns computed over 2021. T-statistics are shown in parentheses. Coefficients statistically-significant at 90% level are shown in bold; *** indicates 99% statistical significance.

| | E[R] 2021 | Std[R] 2021 | Skew[R] 2021 | Kurt[R] 2021 |
|---|---|---|---|---|
| F0 | -9.483e-05 (-0.754) | -0.0011 (-1.152) | -0.1430 (-1.115) | 0.2578 (0.171) |
| F1 | 1.203e-05 (0.052) | **-0.0034 (-1.881)** | -0.1388 (-0.590) | **-6.3461 (-2.287)** |
| F2 | -0.0002 (-0.512) | **0.0061 (2.142)** | **0.6491 (1.738)** | 6.6953 (1.520) |
| **Intercept** | **0.0011 (3.622)** | **0.0273 (12.035)** | 0.4660 (1.570) | **5.9926 (1.711)** |

| Model Adj. R2 | -1.8% | 4.5% | 0.8% | 3.4% |
|---|---|---|---|---|

TABLE 13. Summary of regression results of various return characteristics on optimal approximate factors of the actual count of mentions of ESG terms by corporations in the SEC filings over 2019-20. The relevant ESG keywords were proposed by SASB. The regressions are conducted contemporaneously on the 2019-20 data. T-statistics are shown in parentheses. Coefficients statistically-significant at 90% level are shown in bold; *** indicates 99% statistical significance.

| | E[R] 2019/20 | Std[R] 2019/20 | Skew[R] 2019/20 | Kurt[R] 2019/20 |
|---|---|---|---|---|
| F0 | **-3.039e-06 (-1.952)** | **-3.656e-05 (-2.122)** | -0.0009 (-0.584) | -0.0060 (-0.239) |
| F1 | **-2.74e-06 (-2.169)** | -1.302e-05 (-0.926) | -0.0002 (-0.127) | 0.0082 (0.398) |
| F2 | -6.985e-06 (-1.580) | -5.606e-05 (-1.140) | -0.0042 (-0.971) | -0.0258 (-0.360) |
| Intercept | **0.0016 (6.306)*** | **0.0396 (13.666)*** | **0.6383 (2.504)** | **16.9207 (4.001)*** |
| Model Adj. R2 | 3.2% | 1.7% | -1.6% | -2.2% |

TABLE 14. Summary of regression results of various return characteristics on optimal approximate factors of the actual count of mentions of ESG terms by corporations in the SEC filings over 2019-20. The relevant ESG keywords were proposed by SASB. The regressions are conducted on forward-looking returns computed over 2021. T-statistics are shown in parentheses. Coefficients statistically-significant at 90% level are shown in bold; *** indicates 99% statistical significance.

| | E[R] 2021 | Std[R] 2021 | Skew[R] 2021 | Kurt[R] 2021 |
|---|---|---|---|---|
| F0 | 3.735e-08 (0.028) | **-2.742e-05 (-2.625)** | -0.0004 (-0.259) | -0.0058 (-0.347) |
| F1 | **1.983e-06 (1.852)** | -1.066e-05 (-1.257) | 0.0003 (0.286) | 0.0007 (0.054) |
| F2 | **6.865e-06 (1.832)** | **-5.67e-05 (-1.912)** | 0.0012 (0.308) | -0.0175 (-0.371) |
| Intercept | **0.0008 (3.695)*** | **0.0283 (16.159)*** | **0.4638 (1.986)** | **9.1309 (3.271)*** |
| Model Adj. R2 | 2.2% | 4.9% | -2.5% | -2.5% |

## 6.3. Discussion of Results

The first optimal approximate factor is the one that explains the most variable within its generating dataset. In our case, these underlying datasets are the existence of mentions of ESG terms (dummy variables) and the actual count of ESG mentions. However, what we see in the regressions of return characteristics on the first three factors is that the first factor is not necessarily relevant to the return characteristics.

In the case of the optimal approximate factors generated from the actual counts of ESG mentions, some of the three factors were statistically significant when used in regressions of return characteristics. For example, the first factor F0 in the count of mentions was 90%+ statistically-significant in explaining variation of contemporaneous returns and their volatility (Table 13). This factor showed a negative relationship with both contemporaneous returns and volatility. Since F0 is negatively correlated with diversity, its negative relationship with the returns shows that corporate focus on diversity has a positive relationship with contemporaneous returns and volatility. The impact of F0 on the forward-looking 2021 returns was negligible, but F0 had a statistically-significant relationship with 2021 volatility (Table 14). Forward looking volatility decreased with F0, which means the volatility increased for firms that focussed on diversity in 2019-20.

The second optimal count-based factor F1 had a 90%+ statistical significance in lowering contemporaneous returns, and also in increasing forward-looking returns (Tables 13 and 14). From Table 10, the second count-based factor had -80% correlation with "hazardous materials" and -64% correlation with "diversity". The increased managerial attention to these subjects is linked with higher contemporaneous and lower future returns. One explanation for this phenomenon may be that corporations choose to focus on popular ESG topics during periods of extra profitability.

The third factor F2 in the counts of mentions dataset was the most interesting: it had over 90% statistical significance in increasing future returns while decreasing the return volatility. This factor had a 97% correlation with "ghg" which stands for "greenhouse gasses". This shows that firms where the management focussed on greenhouse gasses in 2019-20, or at least talked about greenhouse gasses in their corporate filings, achieved higher returns with lower volatility in 2021. In our analysis, we are not able to distinguish between "greenwash" and intentional talk about ESG initiatives.

For the optimal approximate factors of the larger stakeholder population proxied by dummy indicators of the mentions, the first optimal factor F0 was not a significant predictor of returns. Recall that the optimal approximate factors are developed from the data that has no knowledge of returns. Instead, the factors represent the key drivers of variation in the data. Therefore, the optimal factors may or may not contain information pertinent to the return determination.

 

The second optimal factor F0 developed from the dummy variables was negatively related to the contemporaneous average returns, but also negatively related to future volatility and kurtosis. According to Table 9, this factor is negatively correlated with "hazardous materials" (-80%), "product quality" (-61%) and "emissions" (-60%).

The third optimal factor F2 constructed from the dummy variables is associated with increases in average contemporaneous returns, contemporaneous volatility, skewness and kurtosis. In the forward looking returns, the same factor is responsible for increased skewness and volatility. The factor is +58% correlated with product labeling.

Overall, we see that activity in the ESG space is likely to bring higher future volatility of the returns.

# 8. Conclusions

In this paper, we test advanced Big Data and Artificial Intelligence techniques with an application to ESG and contrast these techniques with traditional OLS linear regression analysis. We use the SEC Edgar database to extract a large volume of self-reported ESG-related comments and relate them to the contemporaneous and forward-looking stock returns. We find strong statistical significance in ESG terms predictability of contemporary and future returns characteristics.

Investors potentially consider repeated mentions in the corporate filings something of an acute action item for the corporations, raising corporate risk and, ultimately, stock volatility. The absence of ESG mentions, on the other hand, may signify that the ESG inclusion policy is already successfully implemented and is steadily run by the corporation, raising little concern from the shareholders and the management alike.

On the methodology front, we show a step-by-step application of the AI and Big Data techniques and discuss their findings vis-a-vis linear regression results. .

     

# Appendix I. Correlation of the dummy variables, %.

For each keyword w and for each company c, the dummy variable is:

$d_{wc} = 1$, if *any* of the SEC filings by company c mention keyword w, and

$$d_{wc} = 0 \text{ otherwise} \tag{1}$$

| | greenhouse | emissions | ghg | air quality | wastewater | hazardous materials | human rights | data security | access and affordability | product quality | product safety | selling practices | product labeling | labor practices | employee health | employee safety | diversity | inclusion | employee engagement | business ethics | competitive behavior |
|---|---|---|---|---|---|---|---|---|---|---|---|---|---|---|---|---|---|---|---|---|---|
| greenhouse | 100 | 49 | 19 | 51 | 29 | 24 | -2 | 6 | -3 | 24 | 21 | -3 | 5 | 7 | 26 | 1 | 13 | 15 | 1 | 23 | -3 |
| emissions | 49 | 100 | 21 | 25 | 47 | 37 | 0 | 19 | -4 | 41 | 2 | 21 | 21 | -10 | 6 | 22 | 27 | 20 | 22 | 13 | -4 |
| ghg | 19 | 21 | 100 | 13 | 15 | 39 | 20 | 32 | 6 | 22 | 18 | 6 | 14 | 13 | 16 | 17 | 52 | 78 | 17 | 17 | 6 |
| air quality | 51 | 25 | 13 | 100 | 30 | 14 | -8 | -4 | -2 | 12 | -7 | -2 | -5 | -5 | 11 | -6 | 7 | 14 | 9 | 25 | -2 |
| wastewater | 29 | 47 | 15 | 30 | 100 | 31 | 3 | 33 | -2 | 46 | 5 | 37 | 27 | -6 | 21 | 19 | 7 | 17 | 33 | 19 | -2 |
| hazardous materials | 24 | 37 | 39 | 14 | 31 | 100 | 2 | 31 | 14 | 30 | 20 | -6 | 27 | 14 | 12 | 23 | 24 | 36 | 9 | 16 | -6 |
| human rights | -2 | 0 | 20 | -8 | 3 | 2 | 100 | 16 | -3 | 6 | 9 | -3 | -8 | 20 | 2 | 32 | 4 | 11 | 0 | 22 | 27 |
| data security | 6 | 19 | 32 | -4 | 33 | 31 | 16 | 100 | 15 | 32 | 29 | 15 | 29 | 34 | 13 | 18 | 40 | 34 | 25 | 25 | 15 |
| access and affordability | -3 | -4 | 6 | -2 | -2 | 14 | -3 | 15 | 100 | -4 | -3 | -1 | -2 | -2 | -3 | -3 | 11 | 6 | -3 | -3 | -1 |
| product quality | 24 | 41 | 22 | 12 | 46 | 30 | 6 | 32 | -4 | 100 | 18 | 20 | 30 | 12 | 5 | 20 | 29 | 26 | 20 | 29 | -4 |
| product safety | 21 | 2 | 18 | -7 | 5 | 20 | 9 | 29 | -3 | 18 | 100 | -3 | 34 | 24 | 3 | -9 | 23 | 21 | -9 | 37 | -3 |
| selling practices | -3 | 21 | 6 | -2 | 37 | -6 | -3 | 15 | -1 | 20 | -3 | 100 | 40 | -2 | -3 | -3 | 11 | 6 | 32 | 32 | -1 |
| product labeling | 5 | 21 | 14 | -5 | 27 | 27 | -8 | 29 | -2 | 30 | 34 | 40 | 100 | 14 | 9 | 7 | 19 | 7 | 7 | 22 | -2 |
| labor practices | 7 | -10 | 13 | -5 | -6 | 14 | 20 | 34 | -2 | 12 | 24 | -2 | 14 | 100 | -6 | 25 | 16 | 5 | 9 | 25 | -2 |
| employee health | 26 | 6 | 16 | 11 | 21 | 12 | 2 | 13 | -3 | 5 | 3 | -3 | 9 | -6 | 100 | 4 | -4 | 19 | -8 | 4 | 34 |
| employee safety | 1 | 22 | 17 | -6 | 19 | 23 | 32 | 18 | -3 | 20 | -9 | -3 | 7 | 25 | 4 | 100 | 5 | 15 | 3 | -3 | |
| diversity | 13 | 27 | 52 | 7 | 7 | 24 | 4 | 40 | 11 | 29 | 23 | 11 | 19 | 16 | -4 | 20 | 100 | 51 | 27 | 20 | -8 |
| inclusion | 15 | 20 | 78 | 14 | 17 | 36 | 11 | 34 | 6 | 26 | 21 | 6 | 7 | 5 | 19 | 5 | 51 | 100 | 20 | 20 | 6 |
| employee engagement | 1 | 22 | 17 | 9 | 33 | 9 | 0 | 25 | -3 | 20 | -9 | 32 | 7 | 9 | -8 | 15 | 27 | 20 | 100 | 3 | -3 |
| business ethics | 23 | 13 | 17 | 25 | 19 | 16 | 22 | 25 | -3 | 29 | 37 | 32 | 22 | 25 | 4 | 3 | 20 | 20 | 3 | 100 | -3 |
| competitive behavior | -3 | -4 | 6 | -2 | -2 | -6 | 27 | 15 | -1 | -4 | -3 | -1 | -2 | -2 | 34 | -3 | -8 | 6 | -3 | -3 | 100 |

 

# Appendix II. Correlation of the number of keyword mentions, %.

| | greenhouse | emissions | ghg | air quality | wastewater | hazardous materials | human rights | data security | access and affordability | product quality | product safety | selling practices | product labeling | labor practices | employee health | employee safety | diversity | inclusion | employee engagement | business ethics | competitive behavior |
|---|---|---|---|---|---|---|---|---|---|---|---|---|---|---|---|---|---|---|---|---|---|
| greenhouse | 100 | 79 | -1 | 32 | -1 | 15 | -4 | -4 | -2 | 1 | -2 | 2 | 1 | -2 | 20 | -3 | -3 | 3 | -2 | 0 | -2 |
| emissions | 79 | 100 | 2 | 62 | 4 | 28 | -5 | -3 | -2 | 24 | -5 | 9 | 3 | -5 | 7 | 2 | -3 | 2 | -2 | 0 | -2 |
| ghg | -1 | 2 | 100 | 5 | -3 | 3 | 9 | 7 | 3 | 4 | 7 | 1 | 3 | 32 | 0 | 13 | 3 | 10 | 1 | 11 | 1 |
| air quality | 32 | 62 | 5 | 100 | 1 | 30 | -5 | -6 | -2 | 1 | -4 | -2 | -3 | -3 | 2 | -4 | -2 | 3 | 0 | -1 | -2 |
| wastewater | -1 | 4 | -3 | 1 | 100 | 1 | 54 | 1 | -1 | 29 | -3 | 12 | 1 | -3 | -1 | 11 | -1 | 2 | 4 | 2 | -1 |
| hazardous materials | 15 | 28 | 3 | 30 | 1 | 100 | -6 | 7 | 3 | 10 | -3 | -4 | 40 | -4 | 4 | -5 | 18 | 31 | -4 | -4 | -4 |
| human rights | -4 | -5 | 9 | -5 | 54 | -6 | 100 | 2 | -3 | 10 | 11 | -3 | -5 | 42 | 0 | 20 | -2 | -1 | -2 | 8 | 12 |
| data security | -4 | -3 | 7 | -6 | 1 | 7 | 2 | 100 | 39 | 3 | 10 | 13 | 8 | 5 | -2 | -6 | 0 | 28 | 0 | 4 | 3 |
| access and affordability | -2 | -2 | 3 | -2 | 1 | 3 | -3 | 39 | 100 | -3 | -2 | -1 | -2 | -2 | -2 | 6 | 2 | -2 | -2 | -1 | |
| product quality | 1 | 24 | 4 | 1 | 29 | 10 | 10 | 3 | -3 | 100 | -4 | 41 | 16 | 1 | 0 | 49 | 0 | 20 | 5 | 13 | -3 |
| product safety | -2 | -5 | 7 | -4 | -3 | -3 | 11 | 10 | -2 | -4 | 100 | -2 | 15 | 13 | -4 | -6 | -1 | 11 | -4 | 60 | -2 |
| selling practices | -2 | 9 | 1 | -2 | 12 | -4 | -3 | 13 | -1 | 41 | -2 | 100 | 4 | -2 | -2 | -2 | 0 | 2 | 8 | 32 | -1 |
| product labeling | 1 | 3 | 3 | -3 | 1 | 40 | -5 | 8 | -2 | 16 | 15 | 4 | 100 | -2 | 22 | -3 | 1 | 25 | -2 | 14 | -2 |
| labor practices | -2 | -5 | 32 | -3 | -3 | -4 | 42 | 5 | -2 | 1 | 13 | -2 | -2 | 100 | -5 | 8 | 1 | 5 | 1 | 19 | -2 |
| employee health | 20 | 7 | 0 | 2 | -1 | 4 | 0 | -2 | -2 | 0 | -4 | -2 | 22 | -5 | 100 | -4 | -3 | 0 | -4 | -2 | 52 |
| employee safety | -3 | 2 | 13 | -4 | 11 | -5 | 20 | -6 | -2 | 49 | -6 | -2 | -3 | 8 | -4 | 100 | -2 | 9 | -1 | -3 | -2 |
| diversity | -3 | -3 | 3 | -2 | -1 | 18 | -2 | 0 | -2 | 0 | -1 | 0 | 1 | 1 | -3 | -2 | 100 | 8 | 7 | 1 | -2 |
| inclusion | 3 | 2 | 10 | 3 | 2 | 31 | -1 | 28 | 6 | 20 | 11 | 2 | 25 | 5 | 0 | 9 | 8 | 100 | 34 | 9 | 8 |
| employee engagement | -2 | -2 | 1 | 0 | 4 | -4 | -2 | 0 | -2 | 5 | -4 | 8 | -2 | 1 | -4 | -1 | 7 | 34 | 100 | 0 | -2 |
| business ethics | 0 | 0 | 11 | -1 | 2 | -4 | 8 | 4 | -2 | 13 | 60 | 32 | 14 | 19 | -2 | -3 | 1 | 9 | 0 | 100 | -2 |
| competitive behavior | -2 | -2 | 1 | -2 | -1 | -4 | 12 | 3 | -1 | -3 | -2 | -1 | -2 | -2 | 52 | -2 | -2 | 8 | -2 | -2 | 100 |